\documentclass[11pt]{article}

\usepackage[margin=1in]{geometry}
\usepackage{setspace}
\usepackage[T1]{fontenc}
\usepackage[utf8]{inputenc} % harmless on modern LaTeX; helps older engines

\usepackage{graphicx}
\usepackage{wrapfig}
\usepackage{mdframed}
\usepackage{booktabs}
\usepackage{array}
\usepackage{caption}
\usepackage{subcaption}

\usepackage{amsmath, amssymb}

\usepackage{siunitx}

\usepackage[hidelinks]{hyperref}

\usepackage{multirow}

\usepackage[switch]{lineno}
\usepackage{authblk}

\usepackage[numbers,sort&compress]{natbib}
\usepackage{footmisc}
\AtBeginDocument{}

\usepackage{xcolor}
\definecolor{AnthropicOrange}{HTML}{CC785C} % Anthropic brand accent (warm terracotta)

\definecolor{OpenAIColdGray}{HTML}{080808}
\definecolor{OpenAIWhite}{HTML}{FFFFFF} % cool off-white (paper-friendly)

\definecolor{GrokCharcoal}{HTML}{313131}
\definecolor{GrokOffWhite}{HTML}{F8F7F6}    % warm off-white

\definecolor{PerplexityTurquoise}{HTML}{20808D} % 

\definecolor{InsightTeal}{HTML}{2F7A73}
\definecolor{InsightTealTint}{HTML}{F3FAF9}

\definecolor{ContextSlateBlue}{HTML}{4C6A87}
\definecolor{ContextSlateBlueTint}{HTML}{F4F8FC}

\usepackage[most]{tcolorbox}

\usepackage{enumitem}

\setlist{
  topsep=3pt,        % space above/below the list
  itemsep=1pt,       % space between items
  parsep=0pt,        % space between paragraphs within an item
  partopsep=0pt
}
\usepackage{titlesec}

\titlespacing*{\section}
  {0pt}{0.9\baselineskip}{0.45\baselineskip}

\titlespacing*{\subsection}
  {0pt}{0.7\baselineskip}{0.35\baselineskip}

\titlespacing*{\subsubsection}
  {0pt}{0.55\baselineskip}{0.25\baselineskip}

\titleformat{\paragraph}[runin]
  {\normalfont\normalsize\bfseries}{\theparagraph}{0.75em}{}[]

\titlespacing*{\paragraph}
  {0pt}      % left indent
  {0.25\baselineskip} % space before
  {0.5em}    % space after (because it's run-in, this is horizontal)

\usepackage{etoolbox}
\patchcmd{\thebibliography}{\setlength{\itemsep}{\z@}}{}{}{} % (harmless if not present)
\patchcmd{\thebibliography}{\setlength{\parsep}{\z@}}{}{}{}
\patchcmd{\thebibliography}{\setlength{\parskip}{\z@}}{}{}{}

\AtBeginEnvironment{thebibliography}{%
  \setlength{\itemsep}{0pt}%
  \setlength{\parsep}{0pt}%
  \setlength{\parskip}{0pt}%
}

\title{From Understanding to Creation: A Prerequisite-Free AI Literacy Course with Technical Depth Across Majors}

\author{Amarda Shehu}
\affil{Professor of Computer Science; Vice President and Chief AI Officer\\
George Mason University, Fairfax, Virginia, USA\\
\textit{Correspondence:} \href{mailto:amarda@gmu.edu}{amarda@gmu.edu}}
\date{March 19, 2026} % leave empty to omit date

\begin{document}
%\linenumbers
\maketitle

\begin{abstract}
Most AI literacy courses for non-technical undergraduates emphasize conceptual breadth over technical depth,
relying on non-programming assignments, sandbox tools, and ethical reflection. This paper describes UNIV~182,
a semester-long, prerequisite-free course at George Mason University that teaches undergraduates across majors to
understand, use, evaluate, and build AI systems. The course is organized around five interdependent mechanisms:
(1)~a unifying conceptual pipeline (from problem definition through data, model selection, evaluation, and reflection)
that students traverse repeatedly at increasing technical sophistication; (2)~concurrent integration of ethical
reasoning with the technical progression; (3)~AI~Studios, structured in-class work sessions with documentation
protocols and real-time critique; (4)~a cumulative assessment portfolio in which each assignment builds
competencies required by the next, culminating in a team-based field experiment on chatbot reasoning, subsequently
developed with students into a co-authored research paper, and a final project in which teams build AI-enabled
artifacts and defend them before external evaluators from industry, nonprofits, and education; and (5)~a custom AI
agent that provides structured reinforcement of course concepts outside class hours. The paper situates this design
within a comparative taxonomy of cross-major AI literacy courses, frameworks, and pedagogical traditions. Instructor-coded analysis of student artifacts at four critical assessment stages documents a progression
from descriptive, intuition-based reasoning to technically-grounded design with integrated safeguards, reaching the
Create level of Bloom's revised taxonomy by the final project. To support adoption beyond the originating context, the paper identifies which mechanisms are separable, which require institutional infrastructure, and how the design can be
adapted for settings that range from general AI literacy without technical depth to discipline-embedded offerings. The course is offered as a documented example and resource for instructors, demonstrating that technical depth and
broad accessibility can coexist when the scaffolding is designed to support both.
\end{abstract}

% Optional keywords block (not required by bioRxiv, but often helpful)
\noindent\textbf{Keywords:} {AI Literacy $\cdot$ Undergraduate AI Education $\cdot$ 
Non-Majors $\cdot$ Generative AI $\cdot$ Course Design $\cdot$ 
Scaffolding $\cdot$ Integrated Technical and Ethical Reasoning $\cdot$ 
Bloom's Revised Taxonomy}
% --------------------------------------------------------------------
% Main content (keep your existing section files)
% --------------------------------------------------------------------
\section{Introduction}
\label{sec:intro}

The past three years have seen a rapid expansion of AI literacy efforts aimed 
at undergraduates outside of computer science. Frameworks have been proposed to 
articulate what such literacy should 
encompass~\cite{long-magerko-chi-2020, touretzky2019envisioning, tadimalla-maher-2025, ng-ai-literacy-review-2021}, 
spanning competencies from understanding what AI is, how it works, to evaluating 
its societal implications. Several institutions have translated these frameworks 
into semester-long courses with no prerequisites and have reported measurable 
gains in understanding and 
engagement~\cite{xu2025essentialsailifesociety, Biswas_Fussell_Stone_Patterson_Procko_Sabatini_Xu_2025, Candon_Georgiou_Ramnauth_Cheung_Fincke_Scassellati_2025, Stoyanovich_Lewis_Corbett_Bynum_Rosenblatt_Arif_Khan_2025}. 
Policy bodies have reinforced this direction, calling for AI literacy across 
educational levels~\cite{used-oet-ai-future-2023, unesco-genai-guidance-2023}. 
The result is a growing base of documented curricula, shared materials, and 
empirical feedback.

Courses designed for non-major audiences, however, face a common design 
constraint: students arrive with widely varying preparation, limited time, and 
no necessary commitment to a technical trajectory. The prevailing response has 
been to favor conceptual breadth and ethical awareness over hands-on technical 
practice. Typical assignments rely on non-programming tools, sandbox 
environments, collaborative annotation, and written reflection. The technical 
pipeline, from data to model to evaluation to deployment, is presented in 
lectures and readings but is not something students themselves execute or are 
assessed on. This approach has produced measurable results. It has brought AI 
education to students who would otherwise have no structured exposure to the 
field, and the learning gains documented by multiple institutions are in some 
cases 
substantial~\cite{xu2025essentialsailifesociety, Stoyanovich_Lewis_Corbett_Bynum_Rosenblatt_Arif_Khan_2025, kong2021evaluation}.

Prioritizing breadth carries an underlying cost. Students who learn 
\emph{about} AI without building, probing, or evaluating AI systems may 
struggle to distinguish fluent model outputs from valid 
reasoning~\cite{klingbeil-trust-reliance-2024, shojaee2025illusion}, to 
identify failure modes that surface only through hands-on experimentation, or 
to make informed judgments about AI adoption in their own professional domains. 
The frameworks themselves acknowledge this gap. Work in~\cite{tadimalla-maher-2025} map the higher-order levels of Bloom's 
taxonomy (Analyze, Evaluate, Create) to their curriculum but reserve those 
levels for CS-major instantiations. Work in~\cite{chiu2025_ai_literacy_competency} 
distinguishes literacy (understanding), competency (applying), and fluency 
(creating), noting that most non-major courses operate at the first level. 
Whether these higher-order competencies are accessible to non-majors when the 
course structure is designed to support them currently remains an open question.

\begin{figure*}[htbp]
    \centering
    \begin{tabular}{c}
    \includegraphics[width=0.99\linewidth]{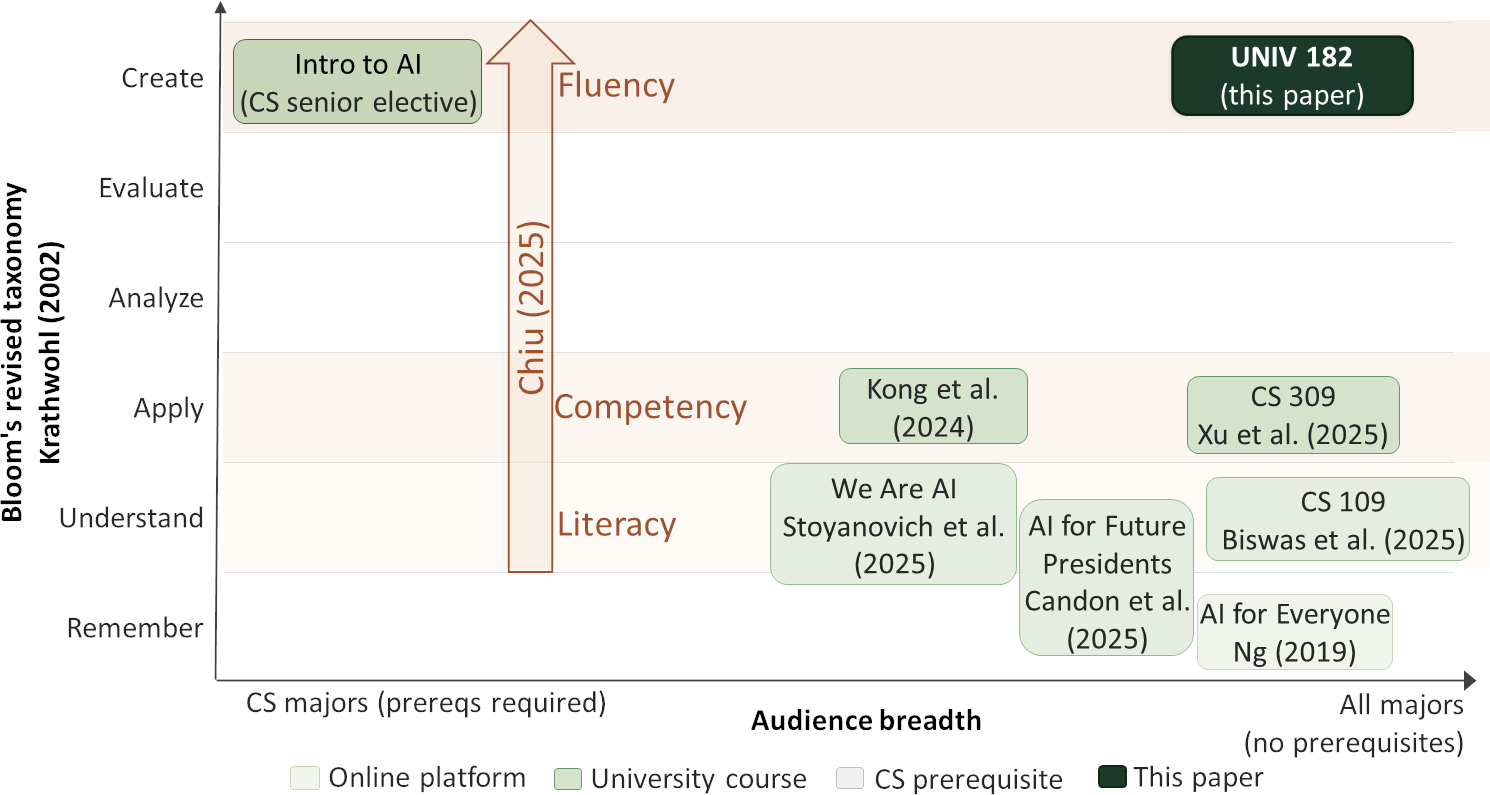}\\[-3mm]
    \end{tabular}
    \caption{Cross-major AI literacy courses positioned along two dimensions: 
audience breadth (horizontal axis) and the highest level of Bloom's revised 
taxonomy~\cite{krathwohl2002revision} that students reach through the 
course's own assessments (vertical axis). Color depth tracks Bloom's level, 
with lighter greens at the Remember and Understand levels deepening to 
forest green at Create. Existing courses for non-majors cluster at the 
Understand and Apply levels. The literacy-competency-fluency 
taxonomy of~\cite{chiu2025_ai_literacy_competency} is shown for reference 
but does not span the full Bloom's progression.}
    \label{fig:landscape}
\end{figure*}

This paper describes a course designed to address that question. UNIV~182 at 
George Mason University is a prerequisite-free, semester-long course designed and offered by the author for 
two consecutive semesters (Fall~2025 and Spring~2026) as part of a 
university-wide AI strategy. Each offering enrolled up to 40 undergraduates 
from diverse majors, with the large majority drawn from non-STEM disciplines 
(demographic details are reported in Section~\ref{sec:design}). The course teaches students to understand and use AI, to evaluate AI systems 
with technical specificity, and to build AI-enabled solutions in a responsible 
manner. This paper uses \emph{literacy} to refer to this full 
progression: the capacity to comprehend what AI systems are, to use them with 
purpose, to evaluate their behavior and limitations with technical specificity, 
and to build with them responsibly under real-world constraints. Ethical 
reasoning is integrated with the technical progression from the first 
assessment to the last, increasing in specificity as students' technical 
knowledge grows. Fig.~\ref{fig:landscape} situates UNIV~182 within the current design space: 
existing courses for non-majors cluster at the Understand and Apply levels of 
Bloom's revised taxonomy, while UNIV~182 occupies the region combining broad 
accessibility with higher-order technical practice.

The course is organized around five interdependent mechanisms, each described 
in detail in Section~\ref{sec:design}:

\begin{enumerate}
\item A \emph{unifying conceptual pipeline} (problem definition, data, 
approach, training/configuration, metrics, evaluation, improvement, reflection) 
that students traverse repeatedly at increasing levels of technical 
sophistication. Students first encounter the pipeline with the model treated as 
a black box, then revisit it for a perceptron, a multi-layer perceptron (MLP), 
a convolutional neural network (CNN) with saliency maps and controlled 
ablations, and finally large language models (LLMs) over the transformer 
architecture. Over the duration of the course, students repeatedly traverse this pipeline with increasing technical 
sophistication.

\item \emph{Concurrent ethical integration.} Rather than confining ethical 
reasoning to a dedicated module, the course embeds it in every assessment. For 
instance, HW1 requires students to analyze ethical implications in a chosen 
societal/economic sector; HW3 requires evaluation of generative AI outputs for 
bias, ownership, privacy, and transparency; and every final project checkpoint 
includes an ethics, privacy, and security review. Two rounds of structured 
debates, held at different points in the semester, make this integration most 
explicit.

\item \emph{AI Studios.} Structured in-class work sessions with 
documentation protocols and real-time critique are scheduled at 
critical points where concepts must become practice before a consequential 
assessment or a leap in technical sophistication. Some prepare students for 
upcoming work (e.g., building classifiers before a CNN interpretation 
assignment, designing experimental protocols before the midterm). Others 
reinforce prior material through hands-on application. Students 
complete work in class, with peers, under observation and guidance from the instructor, which additionally reduces potential cognitive offloading to AI tools.

\item A \emph{cumulative assessment portfolio.} Each assignment builds 
competencies required by the next. Sector briefs (HW1) supply evidence for debates. 
CNN classification and saliency interpretation (HW2) develop experimental discipline 
applied in the midterm. The midterm, a team-based field experiment, establishes that fluent explanation does 
not entail valid reasoning. The midterm findings position students for HW3, which altogether inform the final project, in which teams build AI-enabled artifacts and defend them before external evaluators 
from industry, nonprofits, and education.

\item A \emph{custom AI agent.} This is built by the instructor within the 
university's AI gateway platform and configured with pre-structured 
conversations aligned to individual course topics. Each conversation reinforces 
a concept already introduced in class, presents it in a progression from 
foundational to more demanding formulations, and then poses problems the 
student must solve. The agent extends the scaffolding into hours and settings 
that class time cannot reach, at a pace students control, while requiring them 
to perform the reasoning rather than receive a completed answer.
\end{enumerate}

The contributions of this paper are as follows:

\begin{itemize}
\item The paper describes the first prerequisite-free AI literacy course in 
which non-major undergraduates reach the higher-order levels of Bloom's revised 
taxonomy (Analyze, Evaluate, Create), occupying a region of the design space 
that existing courses and frameworks leave unoccupied (Fig.~\ref{fig:landscape}).

\item The course is presented as a deliberately sequenced design in which the 
five structural mechanisms summarized above function as interdependent 
components. The studios build experimental discipline that assessments require; 
the first debate round establishes a baseline against which the second, held at 
greater technical depth, can be compared; each final project checkpoint draws 
on competencies developed in a specific prior assignment. The contribution lies not in any particular mechanism but in the documented integration of all five, presented at sufficient detail to support 
adaptation at other institutions.

\item A comparative taxonomy of cross-major AI literacy courses, frameworks, 
and pedagogical traditions, organized along five dimensions, situates the 
present work within the existing landscape and provides a reference structure 
for future course design.

\item Instructor-coded analysis of student artifacts across four critical assessment 
stages documents a progression from descriptive, intuition-based 
reasoning to technically-grounded design with integrated safeguards, reaching 
the Create level of Bloom's revised taxonomy by the final project. This 
evidence indicates that the prevailing assumption, that technical depth and 
broad accessibility require fundamentally different course designs, may warrant 
reconsideration when the scaffolding is designed to support both.

\item To support adoption beyond the originating institution, the paper 
identifies which course mechanisms are separable, which require specific 
institutional infrastructure, and how the design can be adapted for settings 
that range from general AI literacy without a technical progression to 
discipline-embedded offerings.
\end{itemize}

The rest of this paper is organized as follows. Section~\ref{sec:related} reviews 
definitions, frameworks, courses, and pedagogical traditions relevant to 
cross-major AI literacy, and positions the present work within that landscape. 
Section~\ref{sec:design} describes the course design, including institutional 
context, student population, topic sequence, and the five pedagogical 
mechanisms. Section~\ref{sec:assessment} details the assessment portfolio and 
the rationale for its cumulative structure. Section~\ref{sec:evidence} presents 
the four-stage analysis of student reasoning and its mapping to Bloom's revised 
taxonomy. Section~\ref{sec:adaptation} discusses adaptation pathways for 
instructors in different institutional and disciplinary contexts. 
Section~\ref{sec:conclusion} addresses limitations and directions for future 
work.
\section{Related Work}
\label{sec:related}

The growing literature on AI literacy for broad audiences is organized below around three threads that converge on the challenge UNIV~182 addresses: cultivating technical depth among non-majors without requiring traditional prerequisites.

\subsection{Definitions and Frameworks}
\label{sec:rw-frameworks}

The term \emph{AI literacy} acquired its most widely cited formulation from Long and Magerko~\cite{long-magerko-chi-2020}, who defined it as ``a set of competencies that enables individuals to critically evaluate AI technologies; communicate and collaborate effectively with AI; and use AI as a tool online, at home, and in the workplace.'' Their work identified 17 competencies organized around what AI is, what AI can do, how AI works, how AI should be used, and how people perceive AI. The AI4K12 initiative by Touretzky et al.~\cite{touretzky2019envisioning} established five ``big ideas'' for K--12 AI education (perception, representation and reasoning, learning from data, natural interaction, and societal impact), shaping much of the subsequent curricular design at the pre-college level. Ng et al.~\cite{ng-ai-literacy-review-2021} offered an exploratory review that organized AI literacy around know and understand AI, use and apply AI, evaluate and create AI, and AI ethics, a taxonomy closely related to the four-quadrant framing adopted in the course described in this paper. More recent reviews~\cite{ng2023review} have documented the rapid proliferation of AI teaching and learning efforts from 2000 to 2020 while noting persistent gaps in post-secondary offerings for non-technical audiences at the time.

Chiu~\cite{chiu2025_ai_literacy_competency} distinguished AI literacy (understanding AI), AI competency (applying AI in professional practice), and AI fluency (creating with AI), arguing that the field conflates these levels to its detriment. Kong, Cheung, and Tsang~\cite{KONG2024100214} developed and evaluated a 14-hour AI literacy course for senior secondary students, using project-based learning to cultivate problem-solving competence and ethical reasoning. Their pre-post design demonstrated gains in students' ability to apply AI concepts to real-life problems, though the course operated at the conceptual and ethical level without requiring students to build or interpret working systems.

The most fully elaborated curricular framework to date is the four-pillar model of Tadimalla and Maher~\cite{tadimalla-maher-2025}. Their pillars span understanding the scope and technical dimensions of AI, learning to interact with generative AI technologies, applying principles of critical, ethical, and responsible AI usage, and analyzing implications of AI on society. They mapped these pillars to Bloom's revised taxonomy and to assessment clusters (knowledge, skills, attitudes, perceptions), producing a comprehensive architecture for course design. However, their framework draws a line: for non-major audiences, the technical pillar emphasizes ``vocabulary building and high-level understanding,'' and the higher-order Bloom's levels (Analyze, Evaluate, Create) are largely reserved for CS-major instantiations. Across nearly all of these frameworks, technical depth and broad accessibility are treated as competing demands, and courses for non-experts resolve the trade-off in favor of breadth.

This trade-off reflects real constraints: time, prerequisite knowledge, heterogeneous preparation, and the risk of alienating students who did not elect a technical path. But it also carries a cost. If AI literacy for non-majors remains at the level of conceptual awareness, ethical reflection, and supervised tool use, it may not equip learners to distinguish fluent outputs from correct ones, to evaluate AI systems under real conditions, or to make informed decisions about deployment of increasingly sophisticated AI technologies (and agents) in their own professional domains. Work in~\cite{klingbeil-trust-reliance-2024} showed experimentally that users overrely on AI advice even when it contradicts available contextual information, and others~\cite{parasuraman-riley-1997} identified patterns of misuse, disuse, and abuse that arise when users lack calibrated understanding of system capabilities.

\subsection{University-Level AI Literacy Courses}
\label{sec:rw-courses}

A growing number of universities have introduced AI literacy courses open to non-majors, and several have been documented in sufficient detail to attempt comparison.

The closest precedent to the work presented in this paper is the sequence at the University of Texas at Austin. Biswas et al.~\cite{Biswas_Fussell_Stone_Patterson_Procko_Sabatini_Xu_2025} described CS~109, a one-credit seminar featuring weekly guest lectures on topics from AI fundamentals to societal implications, open to students, staff, and faculty. Participants reported gains in AI literacy, but the authors identified shortcomings: the rotating lecturers lacked a cohesive narrative, assessment was primarily attendance-based, and readings were often inaccessible to a general audience. Xu et al.~\cite{xu2025essentialsailifesociety} described the subsequent three-credit expansion, CS~309, which adopted a flipped classroom with segmented video lectures, collaborative annotation via Perusall, weekly reflection essays, and five substantive non-programming assignments (algorithms via pseudocode, planning and search via a GridWorld sandbox, probabilistic reasoning via a particle filter simulator, machine learning via Teachable Machine~\cite{TeachableMachine}, and generative AI via Chatbot Arena). The course culminated in an individual ethics project analyzing an AI tool's societal implications. CS~309 achieved strong student satisfaction (mean interest 4.34/5, engagement 4.04/5, $n=180$), and the flipped-classroom model with a single instructor addressed the coherence problems of CS~109. Its defining pedagogical choice is that all assignments are explicitly non-programming, and the culminating assessment is a written analysis rather than a constructed artifact or a defended design.

At Yale, Candon et al.~\cite{Candon_Georgiou_Ramnauth_Cheung_Fincke_Scassellati_2025} designed ``Artificial Intelligence for Future Presidents,'' a course open to all majors that conveys technical information through analogy and demonstration rather than through programming or mathematics. The course aims to produce ``safe, effective, and informed users'' of AI, a framing that centers use and evaluation over construction. Stoyanovich et al.~\cite{Stoyanovich_Lewis_Corbett_Bynum_Rosenblatt_Arif_Khan_2025} developed ``We Are AI: Taking Control of Technology'' at New York University, a public education course using learning circles (small peer-learning groups with trained facilitators) to teach Responsible AI to librarians and professional staff. Pre-post self-assessments showed gains in perceived understanding (from roughly 4/10 to 6/10), and the train-the-trainer model embedded in the learning-circle format demonstrated scalability. The course avoids technical depth, focusing on case studies, ethical matrices, and collaborative discussion. Work in~\cite{kong2021evaluation} described a seven-hour AI literacy course for university students from diverse disciplines in Hong Kong, emphasizing accessible explanations and cross-disciplinary examples. Numerous online courses serve adjacent goals at scale: OpenAI Academy~\cite{OpenAI2025Academy}, Google AI Essentials~\cite{Google2025AIEssentials}, the Digital Education Council's AI Literacy for All~\cite{DEC_AILiteracyForAllCourse}, and Andrew Ng's AI for Everyone~\cite{Ng2019AIforEveryone}, among others. These prioritize conceptual fluency and practical orientation but necessarily sacrifice the structured assessment, studio-based practice, and iterative skill building that a semester-long institutional course can provide.

Table~\ref{tab:landscape} summarizes the landscape along five dimensions that vary independently across these efforts and collectively define the design space for cross-major AI literacy. The pattern is consistent: courses designed for non-expert audiences resolve the breadth-versus-depth trade-off by choosing breadth. The technical pipeline is described but not practiced; students learn \emph{about} AI systems but do not build, probe, or evaluate them. This is a reasonable design choice, and it has produced demonstrable learning gains. It does, however, leave open the question of whether non-majors, given appropriate scaffolding, can reach the higher-order competencies (Analyze, Evaluate, Create) that the field's own frameworks identify as essential.

% ============================================================
% Landscape Table for Related Work section
% ============================================================

\begin{table*}[htbp]
\centering
\caption{Cross-major AI literacy courses and frameworks at the university level. \emph{Bloom's ceiling} indicates the highest level of Bloom's revised taxonomy that non-major students are expected to reach through the course's own assessments. \emph{AI engagement mode} characterizes the dominant relationship students have with AI systems. \emph{Ethics model} distinguishes whether ethical reasoning is delivered as a separate module (parallel), woven into technical content through a shared structure (integrated), or treated as the organizing principle of the course (ethics-centered). The final column identifies what the highest-stakes assessment asks students to produce or perform.}
\label{tab:landscape}
\small
\renewcommand{\arraystretch}{1.25}
\begin{tabular}{p{2.3cm} p{1.6cm} p{1.1cm} p{1.8cm} p{2.2cm} p{1.8cm} p{2.5cm}}
\hline
\textbf{Course / Framework} & \textbf{Institution} & \textbf{Pre\-reqs} & \textbf{Bloom's ceiling (non-majors)} & \textbf{AI engagement mode} & \textbf{Ethics model} & \textbf{Culminating assessment} \\
\hline
\multicolumn{7}{l}{\emph{Frameworks}} \\
\hline
Long \& Magerko \cite{long-magerko-chi-2020} & -- & -- & Unspecified & -- & Included in competencies & -- \\
AI4K12 \cite{touretzky2019envisioning} & -- & -- & Apply & -- & Societal impact (big idea 5) & -- \\
Tadimalla \& Maher \cite{tadimalla-maher-2025} & UNC Charlotte & None & Apply$^{*}$ & Use tools; reflect & Parallel (Pillars 3--4) & AI literacy self-assessment \\
\hline
\multicolumn{7}{l}{\emph{Courses}} \\
\hline
CS 309 \cite{xu2025essentialsailifesociety} & UT Austin & None & Apply & Use sandboxes; solve non-coding problems & Parallel (videos + essays) & Individual written ethics project \\
AI for Future Presidents \cite{Candon_Georgiou_Ramnauth_Cheung_Fincke_Scassellati_2025} & Yale & None & Understand & Observe demos; discuss & Integrated in lectures & Not reported \\
We Are AI \cite{Stoyanovich_Lewis_Corbett_Bynum_Rosenblatt_Arif_Khan_2025} & NYU & None & Understand & Discuss cases; interact with Teachable Machine & Ethics-centered (RAI focus) & Collaborative reflection \\
CS 109 \cite{Biswas_Fussell_Stone_Patterson_Procko_Sabatini_Xu_2025} & UT Austin & None & Understand & Watch lectures; reflect & Parallel (guest lectures) & Weekly reflections \\
Kong et al. \cite{KONG2024100214} & Hong Kong & None$^{\dagger}$ & Apply & Project-based problem solving & Integrated in PBL & Group project (conceptual) \\
AI for Everyone \cite{Ng2019AIforEveryone} & Online (Coursera) & None & Remember & Watch lectures; quiz & Discussed in lectures & Multiple-choice quizzes \\
\hline
\textbf{UNIV 182} (this paper) & George Mason & None & \textbf{Create} & Build classifiers; probe LLMs; construct artifacts & \textbf{Integrated} (shared technical pipeline) & \textbf{Team artifact + public pitch to external evaluators} \\
\hline
\multicolumn{7}{l}{\footnotesize $^{*}$Bloom's levels above Apply are mapped to CS-major instantiations in the Tadimalla \& Maher framework.} \\
\multicolumn{7}{l}{\footnotesize $^{\dagger}$Senior secondary students; no AI-specific prerequisites.} \\
\end{tabular}
\end{table*}

\subsection{Pedagogical Foundations for Technical Depth Without Prerequisites}
\label{sec:rw-pedagogy}

The design of UNIV~182 draws on several pedagogical traditions that support technical depth for mixed-background learners.

\emph{Schema building through repeated traversal.} Donovan and Bransford~\cite{DonovanBransford2005}, synthesizing decades of research on how students learn, argued that robust understanding emerges when learners encounter the same conceptual structures in progressively more complex contexts, building schema that they can adapt and apply. UNIV~182 applies this principle through a stable conceptual pipeline (define the problem, gather data, select an approach, train or configure, choose metrics, evaluate, improve, reflect) that students traverse first in simple contexts (spam filtering, basic classification), then in technically-demanding ones (neural network architectures, saliency interpretation), and finally in ethically-consequential ones (facial recognition, autonomous weapons).

\emph{Studio-based pedagogy}, drawn from design education and architecture, provides a format in which concepts are translated into disciplined practice. Studios are structured work sessions in which students make decisions, document rationale, receive real-time critique, and revise under observation. This format has a long history in design disciplines and has been adapted for computing education, though its use in AI literacy courses is rare, to the best understanding of the author. In UNIV~182, studios serve as the primary site of skill development: students complete work in class, with peers, under protocols requiring evidence and justification.

\emph{Assessment as learning}, rather than assessment of learning, is a third foundation. The literature on formative and portfolio-based assessment~\cite{henderson2011facilitating} emphasizes that assessments designed as discovery tasks, where the process of completing the assessment produces new understanding, can function as the primary pedagogical mechanism rather than as a retrospective check. The cumulative portfolio in UNIV~182, in which each assessment feeds competencies forward into the next, follows this principle. The midterm, for example, is not a test of content recall but a field experiment in which student teams design reasoning probes, test consumer chatbots, and identify for themselves the gap between fluent explanation and valid reasoning, a finding independently documented in the research literature~\cite{shojaee2025illusion, beger2025abstractreasoning, klingbeil-trust-reliance-2024}.

Finally, the course's structured debates draw on \emph{argumentation pedagogy} and the practice of holding opposing claims simultaneously. The framing ``Two Things Can Be True'' requires students to defend positions they may not personally hold, grounding each argument in the shared technical pipeline (problem, data, approach, metric, value judgment). This structure is intended to prevent ethical discussion from proceeding without technical grounding, a failure mode in which positions are asserted from intuition rather than derived from analysis. The integration of technical and ethical reasoning through a single pipeline, rather than the more common pattern of a parallel ethics track, reflects the observation that ethics taught in isolation from technical practice tends to remain abstract~\cite{saltz2019integrating, shapiro2021role-play, burton2015teaching-ai-ethics-scifi}.

Table~\ref{tab:pedagogy} summarizes these four traditions, the design problems each addresses, and the course mechanisms through which each is realized.

\begin{table*}[htbp]
\centering
\caption{Pedagogical foundations informing the design of UNIV~182. Each row identifies a tradition from the learning sciences or professional education, the principle it contributes, the design problem it addresses in cross-major AI literacy instruction, and the course mechanism through which it is realized.}
\label{tab:pedagogy}
\small
\renewcommand{\arraystretch}{1.25}
\begin{tabular}{p{2.0cm} p{3.5cm} p{3.8cm} p{5.6cm}}
\hline
\textbf{Tradition} & \textbf{Core principle} & \textbf{Design problem addressed} & \textbf{Instantiation in UNIV 182} \\
\hline
Schema building through repeated traversal \cite{DonovanBransford2005} 
& Robust understanding develops when learners encounter the same conceptual structure in progressively more complex contexts. 
& Mixed-background students lack a stable frame for organizing new technical material; each topic feels isolated and disposable. 
& A single conceptual pipeline (problem $\rightarrow$ data $\rightarrow$ approach $\rightarrow$ train/configure $\rightarrow$ metrics $\rightarrow$ evaluate $\rightarrow$ improve $\rightarrow$ reflect) is traversed repeatedly, first with simple classifiers, then with neural architectures, then with ethically-consequential deployments. \\
\hline
Studio-based pedagogy (design education) 
& Concepts become disciplined action through structured work sessions with real-time critique, documented decisions, and peer accountability. 
& Non-majors can absorb concepts in lecture but fail to translate them into methodical practice; misconceptions remain latent until high-stakes assessments. 
& AI Studios at key junctures: building classifiers (no-code, then light code via Google Colab~\cite{GoogleColab}), preparing debates, designing experiments, developing prototypes. Protocols require written plans, evidence logs, and rationale before execution. \\
\hline
Assessment as learning \cite{henderson2011facilitating} 
& Assessments designed as discovery tasks produce new understanding rather than measure existing knowledge. 
& Traditional content-verification exams reward recall and penalize the exploratory reasoning that non-majors need to develop; students with weaker technical backgrounds are disproportionately disadvantaged. 
& Cumulative portfolio in which each assessment feeds competencies forward: sector briefs supply evidence for debates; CNN ablations supply experimental discipline for the midterm; the midterm's ``fluency $\neq$ correctness'' finding supplies evaluative habits for the final project. \\
\hline
Argumentation pedagogy; structured discourse 
& Holding opposing claims simultaneously, grounded in shared evidence, develops evaluative judgment that neither lectures nor reflection essays can produce alone. 
& Ethical discussion conducted without technical grounding reduces to opinion exchange; students form positions from headlines rather than from analysis. 
& ``Two Things Can Be True'' debates: teams defend opposing positions on AI deployment (facial recognition, AI art, healthcare, defense), with every argument required to follow the shared technical pipeline (problem $\rightarrow$ data $\rightarrow$ approach $\rightarrow$ metric $\rightarrow$ value judgment). \\
\hline
\end{tabular}
\end{table*}

\subsection{Positioning the Present Work}
\label{sec:rw-positioning}

UNIV~182 occupies a specific position in this landscape. Like CS~309~\cite{xu2025essentialsailifesociety} and ``AI for Future Presidents''~\cite{Candon_Georgiou_Ramnauth_Cheung_Fincke_Scassellati_2025}, it requires no prerequisites and enrolls students across majors. Like the four-pillar framework of Tadimalla and Maher~\cite{tadimalla-maher-2025}, it organizes learning around understanding, using, evaluating, and building AI. It departs from both in requiring non-majors to reach the higher-order competencies that existing courses and frameworks reserve for technically prepared students. Students in UNIV~182 built image classifiers, interpreted CNN saliency maps through controlled ablations, designed and executed reasoning experiments on large language models~\cite{shehu2025consumerchatbotsreasonstudentled}, created with generative AI while evaluating outputs for quality, bias, and provenance, and constructed and publicly defended AI-enabled artifacts before external evaluators from industry, nonprofits, and education. They did so without prior coding experience, without advanced mathematics, and without a CS prerequisite. The claim is not that this approach is superior to the alternatives but that the prevailing assumption, that technical depth and broad accessibility require different course designs, may warrant reconsideration when the scaffolding is designed to support both. The sections that follow describe that scaffolding in detail.
\section{Course Design}
\label{sec:design}

This section describes in greater detail the institutional context, student population, weekly structure, and pedagogical mechanisms of UNIV~182. 

\subsection{Institutional Context and Student Population}
\label{sec:context}

As of the time of the writing of this paper, George Mason University is the largest public university in Virginia, with over $40{,}000$ students. It is a Carnegie R1 institution whose student body is among the most diverse in the United States. The university serves a high proportion of first-generation college students, transfer students, and students from communities historically underrepresented in STEM fields. This demographic context shaped the course design: UNIV~182 could not assume prior exposure to computing, statistics, or formal scientific methodology. The decision to require no STEM prerequisites also reflected the course's role as an entry point in a university-wide AI strategy adopted in 2024 that the the instructor articulated and leads for the university.

The course carries a UNIV prefix, designating it as a university-wide general education offering open to students in any college and any major. General education courses at Mason carry the ``Mason Core'' designation. UNIV~182 is a three-credit course meeting twice weekly in seventy-five-minute sessions. Enrollment was capped at 40 students per offering, with the initial goal of testing efficacy and feasibility before scaling university-wide. Student majors spanned economics, nursing, business, kinesiology, management, policy, computer science, and others, with more than 80\% drawn from non-STEM disciplines. Class standing ranged from first-year to senior. No student was required to have prior coding experience, and the majority had none. In its first offering (Fall~2025), the course also attracted non-traditional adult learners and members of the Northern Virginia community, who either registered for credit or audited.

\subsection{Weekly Structure and Topic Sequence}
\label{sec:weekly}

\begin{figure*}[htbp]
\centering
\begin{tabular}{c}
\includegraphics[width=0.99\textwidth]{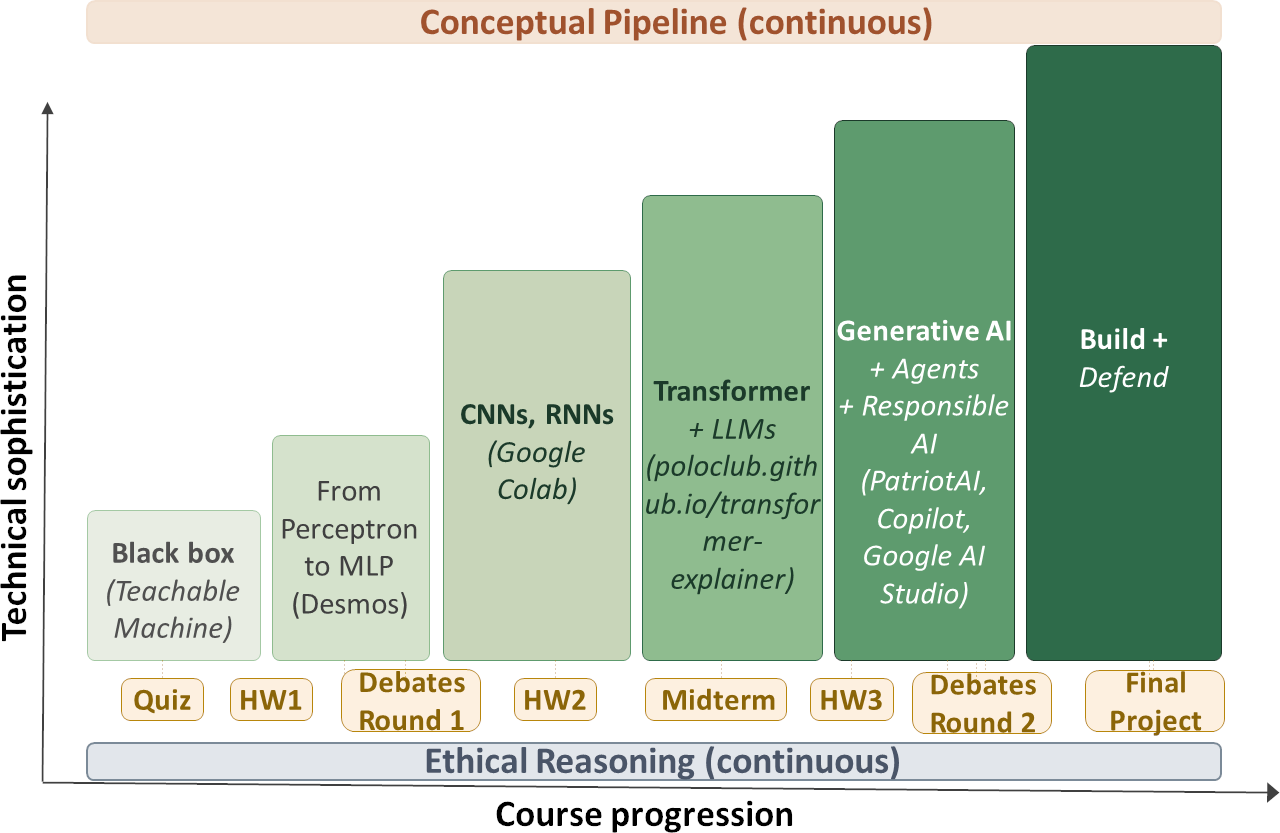}\\[-3mm]
\end{tabular}
\caption{The structure of UNIV~182 as two continuous threads running beneath rising technical sophistication. The conceptual pipeline (orange ribbon) and ethical reasoning (green ribbon) run uninterrupted across the full semester. The blocks represent the technical content students engage with, rising in height to indicate increasing sophistication: from black-box models (Teachable Machine) through perceptrons and MLPs, CNNs and RNNs, transformers and LLMs, generative AI and agents, to the final build-and-defend project. Assessment markers are annotated below. Every assessment intersects both threads: no learning task in the course is purely technical or purely ethical.}
\label{fig:undercurrents}
\end{figure*}

Table~\ref{tab:weekly} presents the week-by-week progression of topics, 
activities, and assessments (Fig.~\ref{fig:undercurrents} visualizes this progression). The sequence enacts the repeated-traversal 
principle described in Section~\ref{sec:rw-pedagogy}: students encounter 
the conceptual pipeline (problem, data, approach, train/configure, metrics, 
evaluate, improve, reflect) at increasing levels of technical sophistication 
across the semester. Ethical reasoning runs concurrently with the technical 
progression from the first assessment onward. HW1 requires students to 
analyze the ethical implications of AI in a specific sector. The structured 
debates (held twice: once during the foundations period and again after 
students have studied transformers and generative AI) require teams to 
defend opposing positions on AI deployment in domains where the stakes are 
high. HW2 requires students to build CNN classifiers, interpret saliency 
maps, and conduct controlled ablations, developing the experimental 
discipline that subsequent assessments demand. The midterm applies that 
discipline to a team-based field experiment testing whether consumer 
chatbots can reason, requiring students to separate answer correctness 
from explanation validity. HW3 requires students to evaluate their own 
generative AI outputs for bias, ownership, privacy, and transparency. The 
final project embeds ethics, privacy, and security review into every 
checkpoint. As students' technical knowledge grows, their ethical reasoning 
gains specificity: early arguments grounded in intuition give way to 
arguments grounded in data provenance, metric selection, and failure mode 
analysis.

% ============================================================
% REVISED TABLE: Linear progression, no week numbers
% ============================================================

\begin{table*}[htbp]
\centering
\caption{Topic sequence, activities, and assessments for UNIV~182 (Fall~2025). The conceptual pipeline and ethical reasoning run concurrently throughout; what increases is the technical sophistication at which both are practiced. Fig.~\ref{fig:undercurrents} visualizes this structure.}
\label{tab:weekly}
\small
\renewcommand{\arraystretch}{1.15}
\begin{tabular}{p{6.7cm} p{6.5cm} p{2.2cm}}
\hline
\textbf{Topics} & \textbf{Activities} & \textbf{Assessments} \\
\hline
\multicolumn{3}{l}{\emph{Foundations: The Pipeline, Data, and First Encounters}} \\
\hline
Introduction to AI; history from Aristotle to Hinton; AI in everyday life & Lectures; in-class discussion and debate comparing early aspirations with realities & -- \\
How AI works: data, learning paradigms (supervised, unsupervised, reinforcement), evaluation & Lectures; data quality and bias case studies & Quiz (in-class) \\
When data and models go wrong & \textbf{Studio}: Hands-on with Teachable Machine; high-profile bias cases & HW1 assigned (AI ethics in a chosen domain) \\
Societal impacts of AI across domains (healthcare, education, climate, policy) & Flipped classroom: small-group discussion seeding HW1 and identifying debate topics & -- \\
\hline
\multicolumn{3}{l}{\emph{Structured Debates, Round 1: Responsible AI Innovation}} \\
\hline
Debate prep: ``Two Things Can Be True'' & \textbf{Studio}: Teams form (silver/red/blue roles); structured worksheet & HW1 due \\
Live debates: facial recognition, AI art, healthcare diagnostics, autonomous defense & Debates with exit tickets; class votes on three strongest teams & Debate performance \\
\hline
\multicolumn{3}{l}{\emph{Deep Learning: From Perceptron to Transformer}} \\
\hline
Perceptron, activation functions, backpropagation, multi-layer perceptrons & Lectures; worked examples building from the simplest network upward & HW2 assigned (CNN + saliency) \\
DNN architectures: CNNs; RNNs & \textbf{Studio}: CNN classifiers in Google Colab; documentation protocols & -- \\
Attention; the Transformer architecture; self-attention & Lectures; from RNN limitations through Word2Vec, GloVe, to Query/Key/Value & HW2 due \\
LLMs and chatbots: from pre-training to policy models & Lecture and discussion: what students interact with vs.\ what is trained & Midterm assigned \\
\textbf{Midterm}: Can chatbots reason? & \textbf{Studio}: Team probe design, controlled execution, dual scoring & Midterm due; teams present \\
\hline
\multicolumn{3}{l}{\emph{Generative AI, Ethics Revisited, and Debates Round 2}} \\
\hline
Generative AI: vibe coding, prompt engineering, agent building & Guest lectures (Google, Microsoft); hands-on sessions & HW3 assigned (Create with AI) \\
Ethics revisited: responsible AI design; security and misuse & Guest lecture (autonomy and security researcher); red-team/blue-team exercises & HW3 due \\
\textbf{Structured Debates, Round 2}: Responsible AI Agent Design & \textbf{Studio}: Teams reform; debates at higher technical specificity & Debate performance \\
\hline
\multicolumn{3}{l}{\emph{Build and Defend}} \\
\hline
Final project: ideation, team formation, prototype development & \textbf{Studio}: Checkpoint 1 (problem, data, method, ethics/risk table) & Checkpoint 1 due \\
Final project: prototype refinement, pitch preparation & \textbf{Studio}: Checkpoint 2 (prototype, updated data plan, safeguards) & Checkpoint 2 due \\
\textbf{Final pitches} to external evaluators & Public presentations + structured cross-examination & Final project due \\
Course wrap-up & Guest reflections; class discussion and feedback on learning journey & -- \\
\hline
\end{tabular}
\end{table*}

% ============================================================
% FIGURE: Pipeline and Ethics as Continuous Threads
% ============================================================

Two features of this structure merit further comment. 

First, the course holds two rounds of structured debates: one at the end of the first quarter of the semester, before students encounter deep learning, and one at the end of the third quarter, after they have studied transformers, LLMs, generative AI, and responsible AI design. The repetition is deliberate. In the first round, students argue about AI deployment using the conceptual pipeline and the evidence from their sector briefs. In the second round, they argue about AI agent design with substantially greater technical specificity, because they can now ground their positions in architectural knowledge, training methodology, and failure mode analysis they did not possess during the first round.

Second, guest lectures from industry and policy practitioners (Google, Microsoft, and a researcher in autonomy, security, and AI ethics) were placed in the generative AI period (final quarter) to bring perspectives complementing those of the course instructor. Students encountered industry tools and policy considerations at the point in the semester when they were working with generative AI systems in HW3, so the guest content is immediately applicable rather than passively received.

\paragraph{Course Materials and Access.} Every lecture was made available in multiple modalities: the original class lecture, a podcast version, and a video version generated through NotebookLM. At a university where access is a stated institutional priority, providing multiple modalities was integral to the design. Each module additionally contained curated current articles selected to connect the topic to contexts students would recognize. For instance, a piece on how students use ChatGPT not to circumvent coursework but to support learning was included in the responsible AI module, giving students a frame for a practice many of them engaged in but had not examined critically. These readings were chosen not for technical depth but for their capacity to link course content to contemporary, real-world experiences.

\subsection{Designing an AI Agent as Course Companion}
\label{sec:agent}

The instructor additionally designed a custom AI agent within the university's PatriotAI platform, an AI gateway that provides access to many LLMs and allows members of the university community to build and experiment in a protected, sandboxed setting. The agent was configured with pre-structured conversations that guided students through specific course topics. Some conversations reinforced foundational concepts: the conceptual pipeline, evaluation methodology, the distinction between training and inference. Others addressed learning paradigms (unsupervised, supervised, and reinforcement learning; prediction versus generation). Others served as preparation for upcoming assessments, walking students through reasoning patterns they would need to apply.

The agent was not a substitute for studio work or instructor feedback but an additional scaffold available outside class hours at a pace students controlled. Its design reflected the same principle that governed the course as a whole: the interaction is structured so that the student performs the reasoning rather than receives a completed answer.

Each pre-structured conversation followed a consistent architecture that other instructors can adapt. The conversation was scoped to a single concept already introduced in class, so the agent reinforced rather than introduced material. Content was presented in a deliberate progression from foundational to more demanding formulations of the same idea. The interaction then switched from explanation to active assessment: the agent posed problems requiring the student to apply the concept, structured as multiple-choice questions with differentiated feedback. Correct responses received reinforcement of the underlying reasoning. Incorrect responses received explanation before the agent continued with further questions. The student retained control throughout and could end the session at any point. This structure (progression followed by assessment with formative feedback and student-controlled pacing) is not specific to any platform or model and can be implemented on any system that supports custom agent configuration.

To illustrate, the following template captures the common structure across all agent conversations:

\begin{quote}
\small
\texttt{I am an undergraduate student in [course]. I want to better understand [concept]. The instructor has explained [what was covered in class]. Explain these concepts to me in progression: from simple to medium to higher level of difficulty. Then, prepare questions to test my understanding. Structure them as multiple choice. When I am correct, reinforce why my answer was correct. When incorrect, explain why I was wrong and continue with more questions. When I say `Stop' you stop.}
\end{quote}

\noindent The template encodes four design decisions: the agent is told what the instructor already covered (preventing it from introducing material out of sequence), difficulty is explicitly scaffolded, the interaction moves from reception to active problem-solving, and the feedback loop is asymmetric (correct answers receive reasoning reinforcement; incorrect answers receive explanation before continuation). Instructors adapting this structure would substitute their own course name, concepts, and coverage descriptions.

\subsection{Pedagogical Mechanisms}
\label{sec:mechanisms}

\paragraph{The Conceptual Pipeline.} The eight-stage pipeline (problem, data, approach, train/configure, metrics, evaluate, improve, reflect) serves as the recurring structure of the course. The instructor maintained it as a consistent reference point across the full semester. Students first encounter it with the model treated as a black box, learning to reason about inputs, outputs, and evaluation without needing to understand internal mechanics. They then revisit the same pipeline for a perceptron, where the mechanics become visible for the first time. Subsequent passes apply it to a multi-layer perceptron, then to a convolutional neural network (where evaluation involves saliency maps and controlled ablations), and finally to large language models over the transformer architecture, which represents the course's highest level of technical sophistication. At each stage, the pipeline remains the same, but what changes is the depth at which students reason about each step. By the final project, the pipeline structures the team pitches: teams articulate a problem worth solving, justify their data sources, explain the AI system they built, describe their evaluation plan, and defend their safeguards.

\paragraph{AI Studios.} As referenced earlier, studios are dedicated in-class sessions for active learning, scheduled at key points in the semester where concepts must become practice before a consequential assessment. Some studios prepare students for upcoming work: building classifiers before HW2 on CNN interpretation, designing experimental protocols before the midterm, and developing prototypes through checkpoints before the final pitch. Others reinforce prior material through hands-on application before students are asked to reason independently. In both cases, the format requires students to complete work in the moment, with peers, under observation, which reduces the tendency to delegate effort to AI tools. Each studio follows a protocol: students document their plan before beginning, log decisions and rationale during execution, and participate in a structured debrief afterward. The instructor circulates during sessions to identify confounds, request clarification, and ensure that reasoning is supported by evidence.

To illustrate how studios surface issues that lectures cannot, one example is instructive. When students built their first image classifier to distinguish smiling from non-smiling faces, data collection raised immediate ethical questions. Students declined to upload their own photos, citing consent and provenance. They considered generating synthetic faces, which led to questions about the training data of the generators, traceability to real individuals, and representation across demographic groups. Students found that prompts for image generators required precise specification and that different engines produced distinct patterns. When teams attempted to delegate the task to an AI agent, the agent required extensive guidance, drifted from the assigned task, and failed to complete it. The exercise exposed issues (data collection ethics, tool limitations, prompt engineering as substantive labor, and the gap between marketed capabilities and actual performance) that lectures alone would not have surfaced with comparable immediacy.

\paragraph{Structured Debates: Two Rounds at Different Depths.} The course holds two rounds of structured debates, each using the ``Two Things Can Be True'' framing. In each round, teams adopt assigned roles: a silver moderator who frames the question and manages the session, a blue team that advocates for adopting an AI technology, and a red team that assesses individual, community, and societal risks. Each argument must follow the shared technical pipeline (Problem $\rightarrow$ Data $\rightarrow$ Approach $\rightarrow$ Metric $\rightarrow$ Value Judgment), which requires ethical claims to be grounded in technical specifics. Structured worksheets guide preparation, and exit tickets collected after each debate document what students identified as the strongest argument from the opposing side, providing a record of perspective-taking. Round~1 addresses AI deployment across consequential sectors, such as facial recognition, AI-generated art, healthcare diagnostics, and autonomous defense systems. Students draw on their HW1 sector briefs and the conceptual pipeline at the level of data and evaluation they have learned to that point. Round~2, held after students have studied transformers, LLMs, generative AI tools, and responsible AI design through both lectures and guest practitioners, addresses responsible AI agent design. 

\paragraph{Scaffolding from No-Code to Light Code.} The progression from Teachable Machine~\cite{TeachableMachine} to Google Colab~\cite{GoogleColab} notebooks was deliberate. Students who had never written code began by training classifiers through a graphical interface, developing intuitions about training data, class boundaries, and overfitting before encountering any syntax. When students moved to Colab notebooks, the notebooks were pre-structured with explanatory comments, and the coding required was light: modifying parameters, running cells, interpreting outputs, and documenting observations. The goal was not to teach programming but to give students sufficient command of the tools to perform controlled experiments, interpret model behavior with specificity, and learn by doing.

\paragraph{Teaching Technical Depth Through Staged Construction.} By mid-semester, students had internalized a guiding principle of the course: every new architecture is motivated by the limitations of what came before. Teaching the transformer~\cite{vaswani2017attention} accordingly began not with the architecture itself but with what earlier models could not solve. RNNs~\cite{SHERSTINSKY2020132306} processed sequences but lost long-range dependencies. Word2Vec~\cite{NIPS2013_9aa42b31} captured word relationships but produced static representations. GloVe~\cite{pennington-etal-2014-glove} incorporated global statistics but could not handle context-dependent meaning. Each limitation was presented as a design constraint, and the next architecture was presented as a response to it. Query, Key, and Value matrices~\cite{vaswani2017attention} were introduced as a mechanism by which each token determines which neighboring tokens are most relevant to its meaning, and multi-head attention was presented as parallel computations capturing different relational structures. The latter connected naturally to material students already knew: in the CNN module, they had learned that multiple kernels capture different patterns in an image. Because each step was motivated by a problem students had just encountered, the progression remained accessible to students without CS backgrounds. Section~\ref{sec:evidence} presents evidence bearing on this claim.
% ============================================================

\section{Assessment Portfolio}
\label{sec:assessment}

The assessment structure of UNIV~182 is a cumulative portfolio in which each assignment builds the competencies required by the next. Two principles guided the design: every assessment should function as a discovery task (producing new understanding, not merely measuring existing knowledge), and the portfolio as a whole should require breadth while providing multiple entry points for students with different strengths. Table~\ref{tab:assessment} presents the assessment sequence.

\begin{table*}[htbp]
\centering
\caption{Assessment sequence (chronological). Each assessment builds competencies that feed forward into subsequent assignments. Bloom's level: Un = Understand, Ap = Apply, An = Analyze, Ev = Evaluate, Cr = Create.}
\label{tab:assessment}
\small
\renewcommand{\arraystretch}{1.2}
\begin{tabular}{p{3.5cm} p{5.8cm} p{6.3cm}}
\hline
\textbf{Assessment} & \textbf{Bloom's Level + Scaffold} & \textbf{Feed-Forward Function} \\
\hline
HW1: Sector brief (individual) & Un+An: Map case to pipeline; convert intuition to referenced claim & Evidence pool for debates; team recombination \\
Debates: ``Two Things True'' (team, live) & Ev: Studio-prepped argument via shared pipeline across high-stakes deployments & Normalizes complexity of real-world evaluation; reinforces evidence-based reasoning \\
HW2: CNN + saliency (individual, notebook) & Ap+An: Studio classifiers $\rightarrow$ mechanistic accountability via controlled ablations, saliency interpretation & Experimental discipline (logs, controlled edits) transfers to midterm protocols \\
Midterm: Chatbot reasoning (team) & An+Ev: Design probes; controlled comparison; score correctness $\neq$ explanation separately & ``Fluency $\neq$ correctness'' finding; evaluative habits for HW3 and final safeguards \\
HW3: Create with AI (individual) & Ap$\rightarrow$Ev: Create artifact, log process, evaluate quality/bias/ownership/privacy/transparency & Ethics as experienced constraints; risk categories carry into final \\
Final: AI artifact + pitch (team, external) & Cr+Ev: Checkpoint studios $\rightarrow$ timed pitch to external evaluators; defend problem, data, design, safeguards & Transfer test: method and accountability under scrutiny by evaluators external to the course \\
\hline
\end{tabular}
\end{table*}

\subsection{The Midterm: A Field Experiment on Chatbot Reasoning}
\label{sec:midterm}

The midterm is designed not as a content-recall test but as a team-based field experiment. Each team selects a reasoning domain (mathematical, causal, spatial, or analogical), designs a set of probes, administers them to multiple consumer chatbots under controlled conditions, and scores answer correctness and explanation validity as separate dimensions. The explicit separation of these two dimensions is intentional: studies have shown that novice users routinely treat fluent explanation as evidence of correct reasoning~\cite{klingbeil-trust-reliance-2024, shojaee2025illusion, beger2025abstractreasoning}. Teams found that models could produce correct answers accompanied by invalid reasoning, or plausible explanations for incorrect answers. They documented inter-rater disagreement and observed that ambiguity in evaluation is a characteristic of working with AI systems under real conditions, not a flaw in the assessment design. The Fall~2025 midterm results were subsequently developed into a co-authored research paper~\cite{shehu2025consumerchatbotsreasonstudentled}, providing additional evidence that non-major undergraduates can contribute to publishable AI research when the course structure supports them. An IRB exemption was obtained to permit dissemination.

\subsection{The Final Project: Public Defense Before External Evaluators}
\label{sec:final}

The final assessment requires teams to defend AI-enabled artifacts before evaluators external to the course. Teams are framed as start-ups or nonprofits building AI-enabled solutions. Through checkpoint-driven studios, they develop a problem statement, justify data sources, explain the AI system, describe evaluation plans and safeguards, and prepare structured pitches. In the final week, teams present to invited external evaluators from industry, nonprofits, and education, who conduct structured cross-examination on feasibility, risk, and responsibility.

In Fall~2025, six teams presented projects spanning energy efficiency, retail loss prevention, misinformation in advertising, behavioral intervention, and transportation safety. Each team articulated a problem, justified data sources, explained the system, described evaluation plans and safeguards, and responded to questions probing feasibility, risk, and responsibility. Evaluator feedback indicated that students were reasoning about impact at a level the evaluators considered comparable to or exceeding what they typically observe in more advanced settings. The specificity of student responses reflected not individual aptitude but the cumulative effect of the course's scaffolding: by the time of the pitch, each team had already refined its problem statement, data plan, and safeguards through two checkpoint studios under instructor critique.

The external evaluators were selected to represent distinct professional perspectives (industry, nonprofit, education) and were briefed on the course's objectives and student backgrounds before the session. Their role was not to assign grades but to pose the kinds of questions professionals face when proposing AI-enabled solutions: whether the problem is real, whether the data is adequate, what could go wrong, and who is affected. The presence of evaluators external to the course transformed the final assessment into a transfer test, requiring students to defend their work before people who had no prior familiarity with it, as well as positioning students to present a portfolio of their work in front of potential employers.
\section{Observations and Evidence}
\label{sec:evidence}

The observations and evidence reported in this section draw largely from the Fall~2025 offering of UNIV~182.

\subsection{Completion and Engagement}
\label{sec:completion}

Of the 38 students enrolled in Fall~2025, 33 completed the course with a passing grade, yielding a completion rate of 87\%. The five students who did not pass submitted between one and five of the eight assessed components, a pattern consistent with disengagement rather than sustained effort followed by failure. No student who submitted all eight assessments received a failing grade. The cumulative portfolio structure, in which each assessment builds competencies required by the next, is designed to produce mastery among students who remain engaged; the primary failure mode observed was departure from the course rather than inability to meet its demands.

\begin{figure}[htbp]
\centering
\includegraphics[width=\linewidth]{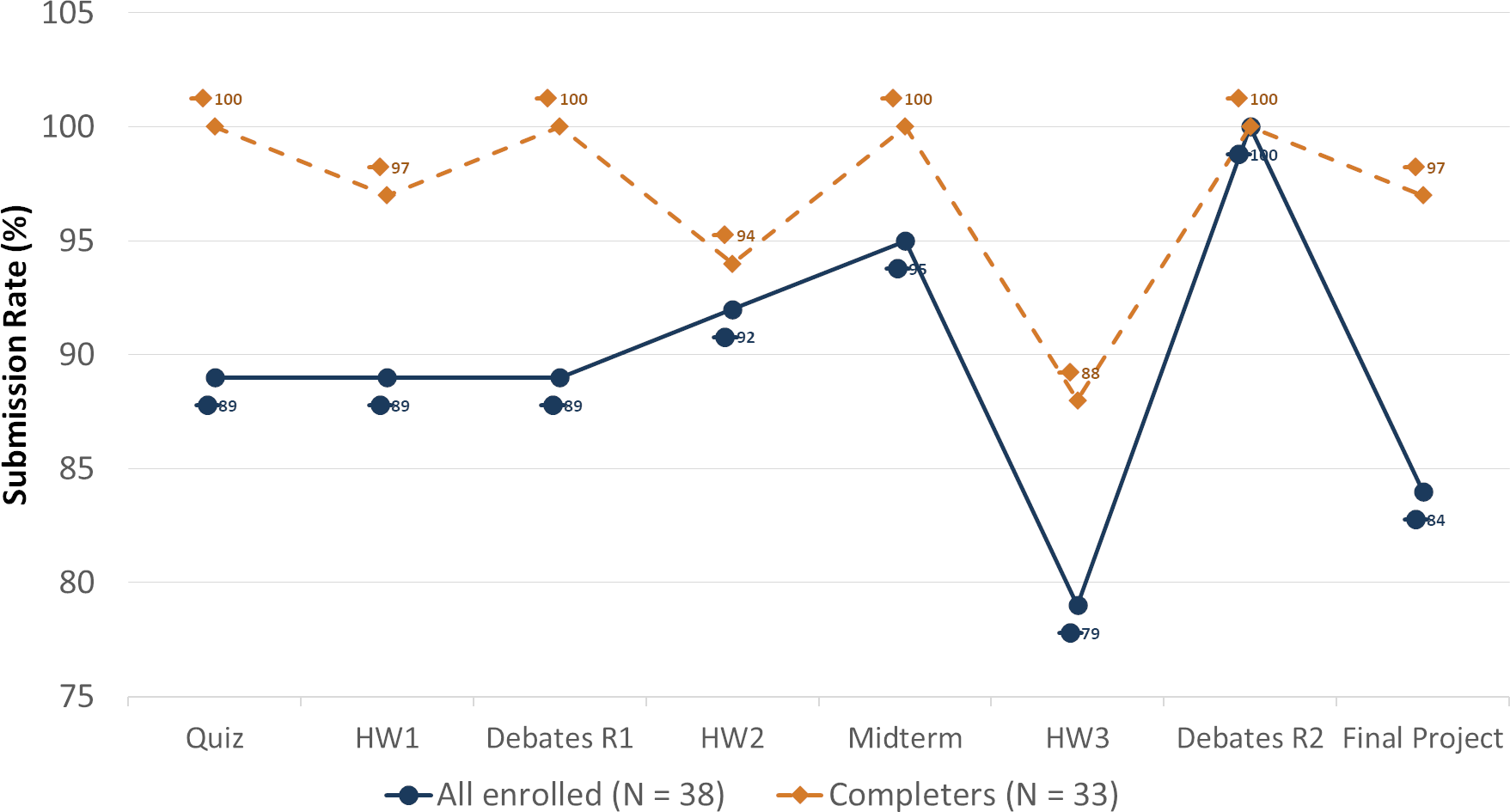}
\caption{Assessment submission rates across the semester for all enrolled students (solid line, $N = 38$) and course completers (dashed line, $N = 33$). Assessments are ordered chronologically. For all enrolled students, submission rates held steady near 89\% through the first three assessments, rose to 95\% at the midterm, dipped to 79\% at HW3 (an individual generative AI assignment in the final quarter), recovered to 100\% for Debates Round~2 (an in-class team activity), and settled at 84\% for the final project. The completers line shows that among students who finished the course, participation remained above 88\% at every assessment point, with the same HW3 trough and Debates Round~2 recovery visible at a smaller magnitude.}
\label{fig:submission_rates}
\end{figure}

Fig.~\ref{fig:submission_rates} tracks submission rates for all 38\footnote{while the course started at capacity, due to visa issues with international students the final enrollment stood at $38$ students.} enrolled students across the eight assessed components in chronological order. Submission rates held steady at approximately 89\% through the first three assessments (quiz, HW1, Debates Round~1), rose to 92\% for HW2 and 95\% for the midterm, dipped to 79\% for HW3 (the individual generative AI assignment in the final quarter), recovered to 100\% for Debates Round~2 (an in-class team activity), and settled at 84\% for the final project. Two observations are notable. First, the midterm's peak participation, at a point where the course's technical demands were highest, suggests that the studio preparation (designing experimental protocols under instructor observation) lowered the barrier to an otherwise demanding task. Second, the dip at HW3 and partial recovery for the final project is consistent with broader findings that team-based and in-class formats sustain engagement more effectively than individual out-of-class assignments, in part because peer accountability reduces the effort reduction that work in~\cite{karau1993social} terms \emph{social loafing} (see also work in~\cite{aggarwal2008social}).

% ============================================================
% FIGURE: Submission Rates (fig_submission_rates)
% ============================================================

The midterm required teams to design reasoning probes, administer them to consumer chatbots under controlled conditions, and produce a written analysis separating answer correctness from explanation validity. Despite being the most technically-demanding assessment to that point in the semester, it achieved the highest submission rate of any component (95\%, 36 of 38 students). This is consistent with the scaffolding claim: the preceding studios and assignments prepared students for a level of experimental discipline that would have been difficult to reach without them. The midterm results, including the proportion of chatbot responses exhibiting correct answers accompanied by invalid reasoning, are reported in the co-authored paper~\cite{shehu2025consumerchatbotsreasonstudentled}. The midterm also functioned as direct instruction in experimental design: students had to formulate hypotheses, select controlled variables, and design evaluation protocols before running their experiments.

%Debates Round~2 was scored on a completion basis and all 33 completers participated. The numerical scores therefore do not capture the shift in reasoning quality between the two debate rounds. That evidence is reported in Section~\ref{sec:progression}, where a content analysis of debate preparation worksheets and exit tickets documents the transition from general societal concern toward technically grounded argumentation.

%Completion and submission rates establish that students remained engaged across a semester of increasing technical demands. That evidence requires analysis of what students produced. Section~\ref{sec:progression} presents evidence that students reached higher-order competencies, based on instructor coding of student artifacts at four key assessment stages.

Institutional course evaluations administered at the end of the Fall~2025 semester provide a complementary source of evidence. Of the respondents, 75\% were freshmen, and 25\% were seniors; 60\% reported taking the course as a Mason Core IT \& Computing (general education); the rest reported taking it as an elective. Table~\ref{tab:evals} reports student evaluations. Three items received the highest mean ratings (4.58/5.00 each, all with 
medians of 5.00): the learning environment facilitating engagement with course content, the use of technologies and tools increasing engagement, and the encouragement of diverse perspectives. The first two recorded a minimum respondent score of~4, indicating that no student who responded rated them below ``agree.'' For a course requiring non-majors to build classifiers, interpret saliency maps, and probe LLMs, the ceiling concentration on engagement items is notable. The item on diverse perspectives is consistent with the debate structure's emphasis on holding opposing positions simultaneously. Self-reported understanding of main concepts was also high (4.42~$\pm$~0.51, median~4.00, min~4), and students rated the variety of learning opportunities positively (4.45~$\pm$~0.69, median~5.00).

For context, Xu et al.~\cite{xu2025essentialsailifesociety} reported mean interest of 4.34/5 and mean engagement of 4.04/5 for CS~309 ($n = 180$), a course that is explicitly non-programming and culminates in a written analysis rather than a constructed artifact. The engagement ratings reported here (4.58 for learning environment, 4.58 for technology  use) were obtained from a course with substantially greater technical demands. The instruments differ, and the comparison is not controlled, but the pattern suggests that technical depth did not diminish perceived engagement. These self-report measures capture students' experience of the course rather than demonstrated competency; the latter is addressed through the artifact-based analysis in Section~\ref{sec:evidence}. Open-ended responses on the same instrument corroborated several design elements. Students identified the structured debates, the midterm, and the final project as the most valuable aspects of the course. 

%One respondent described the value of ``being able to figure out what works, and what doesn't'' without the pressure of producing a single correct answer, a characterization that aligns with the assessment-as-learning principle underlying the portfolio design. 

\begin{table}[htbp]
\centering
\small
\caption{Selected items from institutional course evaluations, Fall~2025. All items 
scored on a 1--5 Likert scale. Items were selected for relevance to the 
course design claims; the full instrument contained 14~items.}
\label{tab:evals}
\begin{tabular}{p{8cm} c c c c}
\toprule
\textbf{Item} & \textbf{Mean} & \textbf{Median} & \textbf{SD} & \textbf{Min} \\
\midrule
Environment facilitated engagement 
  with content & 4.58 & 5.00 & 0.51 & 4 \\
Technologies/tools increased 
  engagement & 4.58 & 5.00 & 0.51 & 4 \\
Encouraged expression of diverse 
  perspectives & 4.58 & 5.00 & 0.67 & 3 \\
Learned through variety of 
  opportunities & 4.45 & 5.00 & 0.69 & 3 \\
Gained understanding of main 
  concepts & 4.42 & 4.00 & 0.51 & 4 \\
\bottomrule
\end{tabular}
\end{table}

\subsection{Evidence of Reasoning Progression}
\label{sec:progression}

Evidence of reasoning progress in this section is organized as a four-stage trajectory, using instructor-coded analysis of student artifacts to show how reasoning about AI systems evolved from descriptive summary toward technically-grounded, independently-constructed analysis. The analysis culminates with a projection on the Bloom's revised taxonomy.

\subsubsection*{Stage 1: Description (Homework 1)}

\begin{figure}[htbp]
\centering
\includegraphics[width=0.85\textwidth]{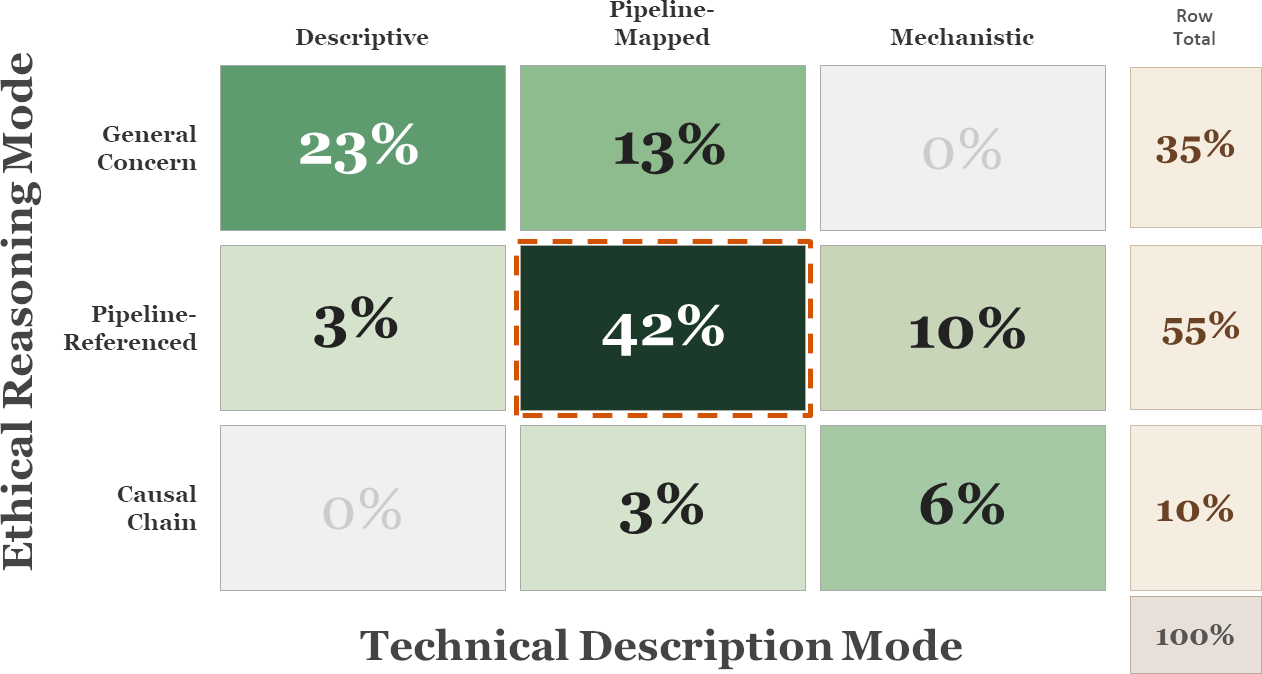}
\caption{Joint distribution of technical description mode and ethical reasoning mode across HW1 submissions. Each submission was coded along two dimensions: how the student described the AI technology (Descriptive, Pipeline-Mapped, or Mechanistic) and how the student reasoned about ethical consequences (General Concern, Pipeline-Referenced, or Causal Chain). The dashed border highlights the modal cell (42\%): students who organized their technical analysis around the information management pipeline but whose ethical reasoning remained at the surface level, referencing the framework without tracing causal paths from design choices to downstream consequences. Only 10\% of submissions contained causal chain reasoning at this stage.}
\label{fig:hw1-coding}
\end{figure}

In the first individual assignment (HW1), students selected an AI technology deployed in an economic/societal sector of their choice and analyzed its implications for information storage, exchange, security, and privacy. Sectors included healthcare, law enforcement, education, finance, defense, creative arts, climate, transportation, and retail surveillance. To characterize the baseline reasoning these submissions represent, the instructor coded each submission along two dimensions: \textit{technical description mode} (how the student described the AI technology) and \textit{ethical reasoning mode} (how the student connected technical features to ethical consequences). Three levels were defined for each dimension, as shown in Fig.~\ref{fig:hw1-coding}.

The results reveal a characteristic gap. The assignment's scaffolding moved most students beyond surface description: 58\% produced pipeline-mapped submissions that organized their analysis around the storage, exchange, security, and privacy framework, and an additional 16\% reached the mechanistic level, explaining how specific technical components (neural networks, data lakes, encryption protocols) function. However, ethical reasoning lagged behind technical description. Across all submissions, 36\% of ethical arguments remained at the level of general concern (claims drawn from intuition, media framing, or broad societal worry without reference to specific technical features). Another 55\% reached the pipeline-referenced level, using the information management framework to organize ethical claims but without tracing a causal path from a design choice to a downstream consequence. Only 10\% of submissions exhibited evidence of \textit{causal chain reasoning}: an argument that traces a specific technical decision (training data composition, storage architecture, exchange protocol) through the pipeline to a specific ethical outcome.

The modal cell in Fig.~\ref{fig:hw1-coding} (pipeline-mapped technical description paired with pipeline-referenced ethical reasoning, containing 42\% of all submissions) captures the baseline beyond which the course was designed to move students. Students at this level could describe AI systems with reasonable specificity and could name ethical concerns in the vocabulary the course provided, but they could not yet derive ethical conclusions from technical premises. Bias, privacy, and accountability were invoked as concerns but were not traced to their origins in data collection, model architecture, or evaluation methodology.

\subsubsection*{Stage 2: Contestation at Foundational Depth (Debates Round 1)}

The first structured debate required teams to defend opposing positions on AI deployment across high-stakes domains (such as facial recognition, AI-generated art, healthcare diagnostics, autonomous defense), with every argument grounded in the shared technical pipeline. The format required students to do something the first homework had not: hold two defensible positions simultaneously and explicitly construct arguments through the pipeline.

The instructor coded each distinct argument advanced across all debate teams into three categories: \textit{general societal concern} (claims drawn from intuition, media framing, or broad societal worry), \textit{pipeline-grounded reasoning} (arguments that trace at least two steps of the pipeline, such as data composition affecting model behavior or evaluation metric choice affecting deployment outcome), and \textit{technically-specific reasoning} (arguments referencing model architecture, training methodology, or observed failure modes at a mechanistic level). 

\begin{figure}[htbp]
\centering
\includegraphics[width=0.75\textwidth]{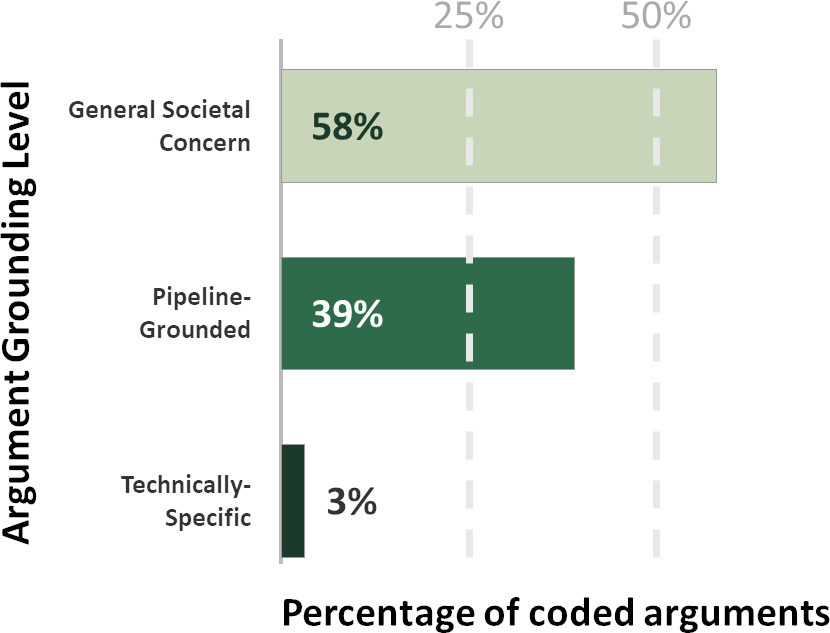}
\caption{Distribution of argument grounding levels across all distinct arguments coded from Debate Round~1. Arguments were classified as general societal concern (claims from intuition or media framing), pipeline-grounded (tracing at least two steps of the technical pipeline), or technically-specific (referencing model architecture, training methodology, or failure modes). Pipeline-grounded reasoning appeared in 39\% of arguments, compared with 10\% of HW1 submissions that contained causal chain reasoning. Technically-specific reasoning was essentially absent, consistent with the course design: the debate preceded the deep learning modules that would supply architectural knowledge.}
\label{fig:debate-r1}
\end{figure}

Fig.~\ref{fig:debate-r1} presents the observed distribution. Compared with HW1, where ethical reasoning was dominated by general concern (35\% of submissions) and surface-level pipeline referencing (55\%), the debate format produced a substantial proportion of pipeline-grounded arguments (39\%). Teams constructed causal chains linking data composition to model behavior to deployment consequence (e.g., tracing how socioeconomic disparities in healthcare data availability could bias diagnostic AI toward wealthier populations, or how historical disciplinary records reflecting racial disparities could propagate through predictive analytics into school resource allocation decisions). Exit tickets collected after the debate documented a related development: students identified the strongest argument from the opposing side and described how their own position had moved, indicating early capacity to simultaneously hold two defensible positions.

The notable absence at this stage was architectural knowledge. Across all coded arguments, technically-specific reasoning (explaining \textit{why} a model misidentifies a face at the level of training distributions, loss functions, or attention mechanisms) was essentially absent (3\%). Arguments about AI failure were asserted rather than derived from an understanding of how models process information. This gap was by design: the debate preceded the deep learning and transformer modules that would supply the architectural grounding that the next assessments would require.

\subsubsection*{Stage 3: Evaluative Engagement Through Creation (Homework 3)}

In the third individual assignment, students used generative AI tools to create an artifact (poem, image, game, application, or short story), documented their iterative process, and critically evaluated the output for quality, bias, ethics, and transparency. This assessment followed the midterm field experiment in which teams had found that consumer chatbots can produce correct answers accompanied by invalid reasoning, a finding that shaped how students approached HW3.

\begin{figure}[htbp]
\centering
\includegraphics[width=0.80\textwidth]{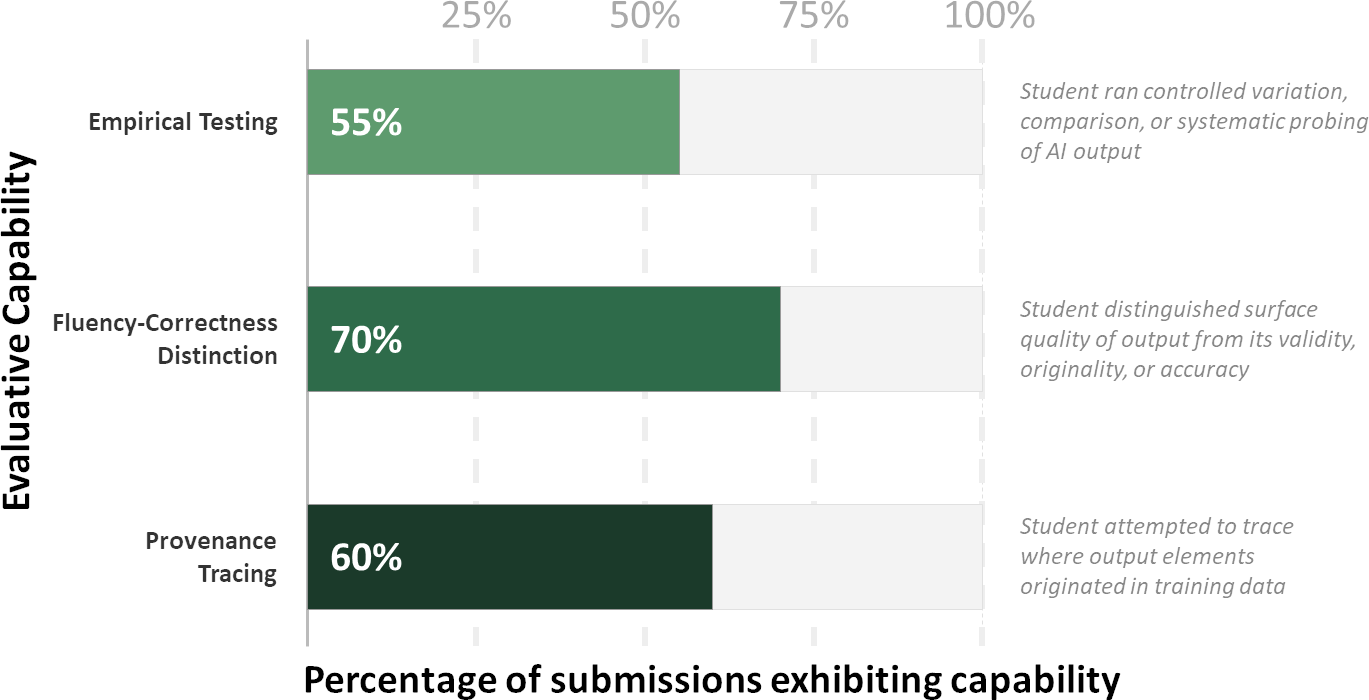}
\caption{Prevalence of three evaluative capabilities across HW3 submissions. Each submission was coded for the presence or absence of empirical testing (controlled variation or systematic probing of AI output), fluency-correctness distinction (separating surface quality from validity or originality), and provenance tracing (attempting to identify where output elements originated in training data or prior works). All three capabilities were essentially absent in HW1 submissions. Their emergence in HW3 is consistent with the scaffolding objective: the AI Studios built experimental discipline, the midterm supplied the fluency-correctness finding, and the cumulative portfolio extended into independent creative evaluation.}
\label{fig:hw3-capabilities}
\end{figure}

To characterize the evaluative reasoning these submissions represent, the instructor coded each submission for the presence or absence of three capabilities that were essentially absent in HW1 and only nascent in Debate Round~1: \textit{empirical testing} (whether the student ran controlled variations, comparisons, or systematic probing of AI output rather than accepting it at face value), \textit{fluency-correctness distinction} (whether the student distinguished the surface quality of AI output from its validity, originality, or accuracy), and \textit{provenance tracing} (whether the student attempted to trace where elements of the output originated in training data or prior works). Fig.~\ref{fig:hw3-capabilities} presents the results.

The change from earlier stages was substantial. More than half of submissions (55\%) exhibited empirical testing: students ran systematic prompt variations to expose color bias in image generation, withheld instructions to test whether an AI could produce novel rather than derivative output, or documented multi-iteration failures to identify the boundaries of what generative tools can and cannot do. Interestingly, these behaviors were not prescribed by the assignment but emerged organically as methodological habits that the AI Studios and the midterm's experimental protocols aimed to develop, and now applied independently by students in a new context.

The fluency-correctness distinction, the midterm's central finding, appeared in 70\% of submissions. Students noted that AI-generated text could sound coherent without being substantive, that outputs defaulting to statistically-probable patterns masked a lack of originality, and that confident system interfaces concealed genuine limitations. The appearance of this evaluative insight in a context different from the one in which it was first established (the midterm's chatbot reasoning probes applied here to individual creative production) constitutes evidence of transfer rather than task-specific performance.

Provenance tracing appeared in 60\% of submissions. Students attempted to identify where AI outputs originated: whether game mechanics were reproduced from identifiable titles, whether poetic forms defaulted to Western canon conventions, whether image generation reflected culturally-specific aesthetics embedded in training data, etc. These inquiries were grounded in specific observations about the artifacts students had produced, connecting the output to the data and processes that generated it.

What remained limited at this stage was architectural explanation. Students could observe that outputs were derivative, that bias existed, and that transparency was lacking, but most could not connect these observations to attention mechanisms, loss functions, or training objectives at a mechanistic level. That architectural grounding is what the deep learning and transformer modules (in the third quarter of the course) were designed to supply, and its presence or absence is what distinguishes the final project stage from HW3.

\subsubsection*{Stage 4: Synthesis Through Design (Final Project)}

\begin{figure}[htbp]
\centering
\includegraphics[width=0.85\textwidth]{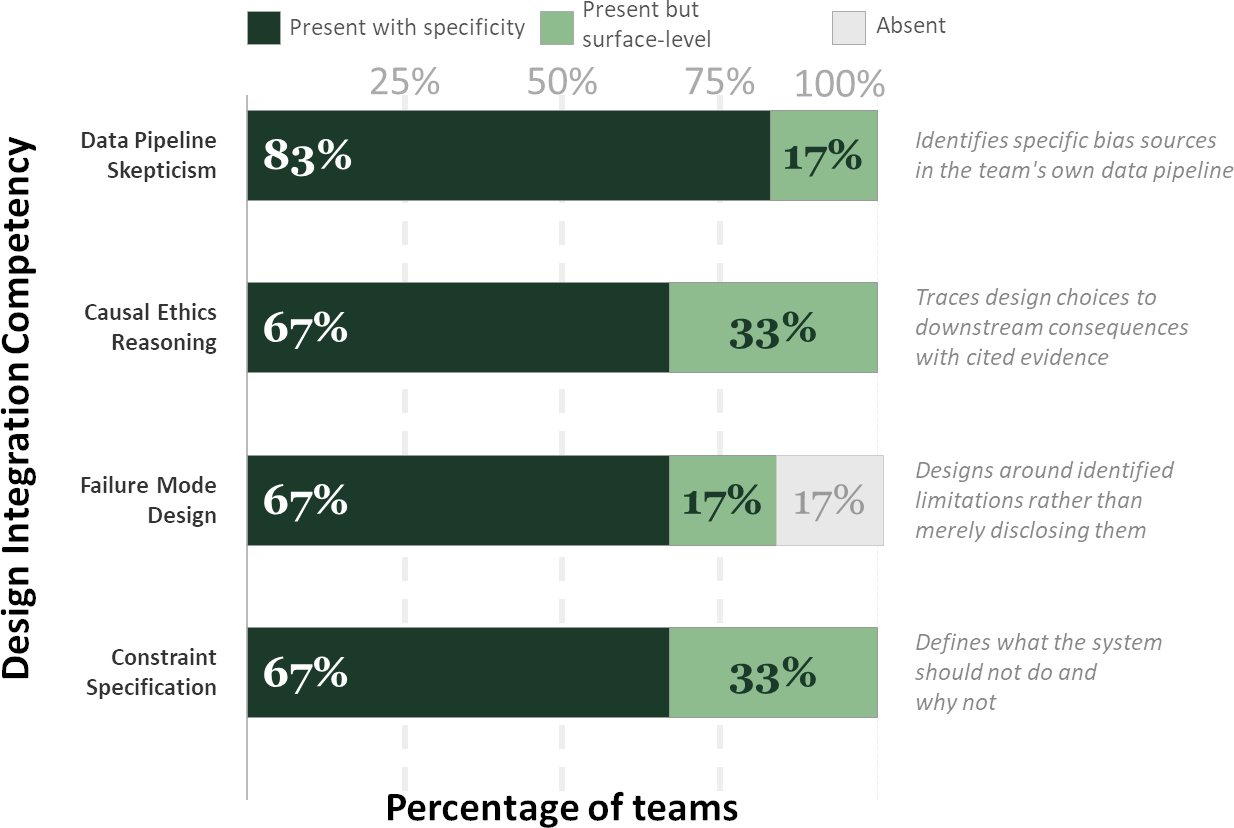}
\caption{Prevalence and depth of four design integration competencies across final project team checkpoint documentation. Each team was coded for data pipeline skepticism (identifying specific bias sources in their own pipeline), causal ethics reasoning (tracing design choices to downstream consequences with evidence), failure mode design (designing around limitations rather than merely disclosing them), and constraint specification (defining what the system should not do). Bars show the proportion of teams exhibiting each competency with specificity (dark), at a surface level (medium), or not at all (light). The near-universal presence of these competencies, which were essentially absent in HW1, is consistent with the claim that the scaffolding mechanisms accumulated across the semester.}
\label{fig:fp-competencies}
\end{figure}

In the final project, teams built AI-enabled artifacts and defended them before invited external evaluators from industry, nonprofits, and education. The checkpoint-driven studio structure (Checkpoints~1 and~2, followed by the public pitch) provided a record of how teams refined their reasoning under iterative critique. To characterize the reasoning these final artifacts represent, the instructor coded each team's checkpoint documentation for the presence and depth of four design integration competencies: \textit{data pipeline skepticism} (whether the team identified specific bias sources in their own data pipeline, traced to particular data sources or processing steps), \textit{causal ethics reasoning} (whether ethical claims traced design choices to downstream consequences with cited evidence rather than listing general concerns), \textit{failure mode design} (whether the team designed around identified limitations rather than merely disclosing them), and \textit{constraint specification} (whether the team defined what the system should not do and articulated why). Each competency was coded at three levels: present with specificity, present but surface-level, or absent. Fig.~\ref{fig:fp-competencies} presents the results.

The contrast with earlier stages is marked. In HW1, only 10\% of the submissions contained causal chain reasoning connecting technical choices to ethical consequences. By the final project, every team exhibited at least surface-level causal ethics reasoning, and two-thirds did so with specificity:tracing how data collection infrastructure shapes the reliability of downstream predictions, how socioeconomic disparities in app usage skew training data, or how decisions made during data collection constrain what a model can and cannot reliably measure. These arguments were grounded in the specific data sources and processing steps of the teams' own projects.

Data pipeline skepticism was the most uniformly developed competency, with 83\% of teams identifying bias sources at the level of specific datasets or processing stages. Teams identified multiple categories of bias in their own pipelines and traced each to a distinct origin. The specificity, classifying bias by origin rather than invoking it as a single undifferentiated concern, represents a qualitative change from the reasoning observed earlier in the course.

Failure mode design appeared with specificity in two-thirds of teams. Rather than listing limitations as caveats, these teams incorporated identified failure modes into their system architecture: setting bounds that the AI could not exceed, deliberately degrading data resolution as a privacy mitigation, planning controlled interference testing before deployment, and requiring human override at specified confidence thresholds. This integration of limitations into the design, rather than their segregation into a risk disclosure section, reflects the kind of principled trade-off analysis that the course's repeated pipeline traversals were designed to develop.

Constraint specification showed a similar pattern. Teams defined what their systems should not do with the same deliberation they applied to what the systems should do: specifying that a behavioral monitoring tool does not constitute medical advice, that predictive classification of individuals requires continuous demographic auditing, or that data collected during emergency operations must be deleted after the mission concludes. These constraints emerged from the teams' own analysis of how their systems could be misused, informed by the cumulative engagement with responsible AI design across the semester.

\subsection{Progression Along Bloom's Taxonomy}

Each of the four assessment stages reported above used a coding scheme specific to that assessment. To translate these stage-specific codes into a common framework, the instructor mapped each coding level to the Bloom's revised taxonomy level it most closely reflects, following the cognitive process definitions established in ~\cite{Conklin2005TaxonomyReview}. The approach adapts the methodology of discipline-specific Bloom's classification tools, most notably the Blooming Biology Tool~\cite{crowe2008biology}, which demonstrated that transparent, domain-adapted rubrics can be applied to student artifacts when the mapping criteria are made explicit. Similar mapping procedures have been used to track cognitive progression across assessment sequences in undergraduate science courses~\cite{momsen2010just, zaidi2017climbing}. 

\begin{figure}[htbp]
\centering
\includegraphics[width=0.85\textwidth]{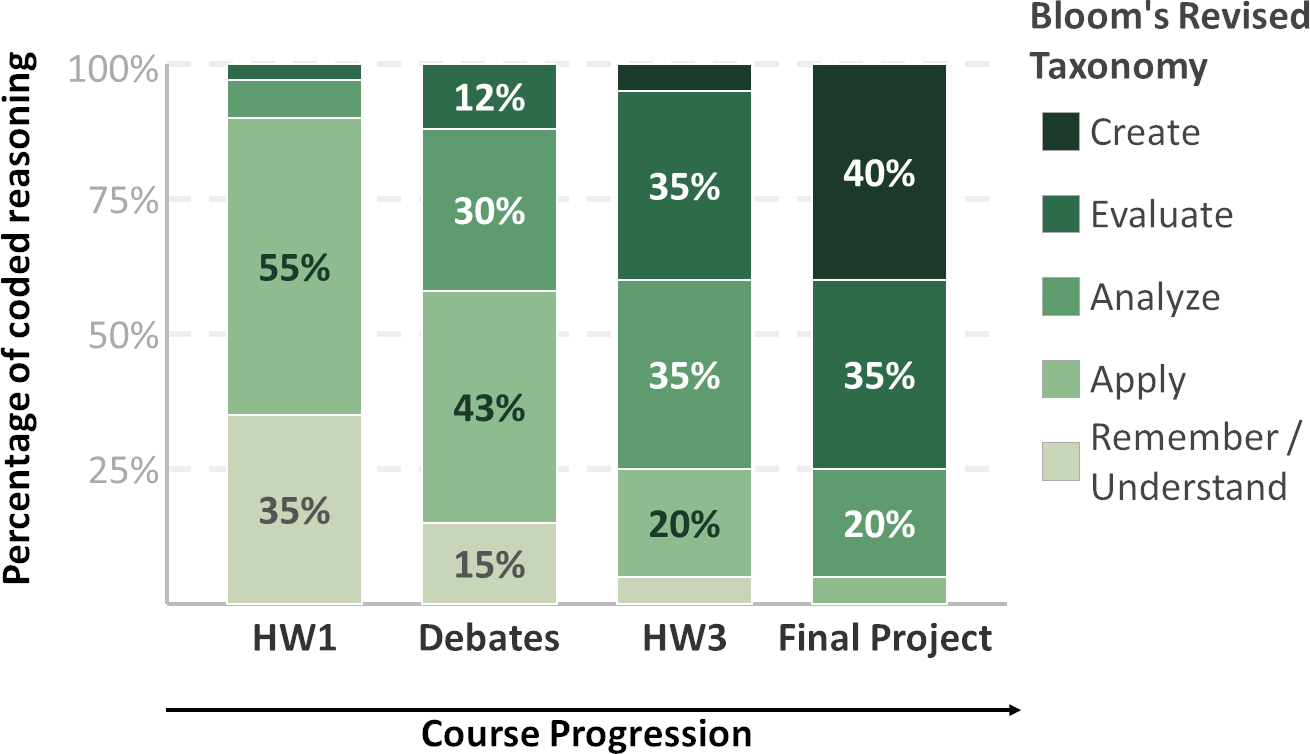}
\caption{Distribution of coded reasoning across Bloom's revised taxonomy levels at four assessment stages. At each stage, the proportion of reasoning at each taxonomic level was estimated from the stage-specific codings reported in the preceding subsections. In HW1, reasoning was concentrated at Remember/Understand and Apply (90\% combined). By the final project, 95\% of coded reasoning operated at the Analyze, Evaluate, or Create levels, with Create appearing at substantial scale (40\%) for the first time. The upward trajectory of the distribution across the semester is consistent with the claim that the course's scaffolding mechanisms produced cumulative growth in higher-order competencies among non-major undergraduates.}
\label{fig:blooms-progression}
\end{figure}

The mapping criteria were defined before coding, and each mapping is reported with its justification in Table~\ref{tab:blooms-mapping}. Where a single submission or argument exhibited reasoning at multiple levels, it was assigned to the highest level observed. The resulting distributions are shown in Fig.~\ref{fig:blooms-progression}.

\begin{table}[htbp]
\centering
\caption{Mapping of assessment-specific codes to Bloom's revised taxonomy levels. Each row shows the assessment stage, the code assigned during stage-specific analysis, the Bloom's level to which the code was mapped, and the cognitive process that justifies the mapping. Cognitive process definitions follow from work in~\cite{Conklin2005TaxonomyReview}: \emph{Remember/Understand} involves retrieving or constructing meaning from information; \emph{Apply} involves carrying out a procedure in a given situation; \emph{Analyze} involves breaking material into parts and determining how parts relate to one another; \emph{Evaluate} involves making judgments based on criteria and standards; \emph{Create} involves putting elements together to form a novel, coherent whole.}
\label{tab:blooms-mapping}
\small
\begin{tabular}{p{1.5cm} p{4.2cm} p{2cm} p{6.5cm}}
\toprule
\textbf{Assessment} & \textbf{Stage-specific code} & \textbf{Bloom's level} & \textbf{Justification (cognitive process)} \\
\midrule
\multirow{3}{2.2cm}{HW1}
& Descriptive (technical) or General Concern (ethical)
& Remember / Understand
& Recall or surface-level characterization of AI systems; no decomposition or application of a framework \\[6pt]
& Pipeline-Mapped (technical) + Pipeline-Referenced (ethical)
& Apply
& Use of a provided framework (the information management pipeline) to organize analysis, without independent decomposition \\[6pt]
& Mechanistic (technical) or Causal Chain (ethical)
& Analyze
& Tracing relationships between technical components and downstream consequences; differentiating causes from effects \\
\midrule
\multirow{3}{2.2cm}{Debate Round~1}
& General societal concern
& Apply
& Constructing claims from intuition or media framing; assembling an argument without decomposing it through the pipeline \\[6pt]
& Pipeline-grounded
& Analyze
& Tracing at least two steps of the technical pipeline to derive a conclusion; differentiating data, model, and outcome \\[6pt]
& Technically-specific
& Evaluate
& Referencing architecture or training methodology to judge system behavior against an independent technical criterion \\
\midrule
\multirow{3}{2.2cm}{HW3}
& No evaluative capabilities present
& Apply
& Using generative AI tools and documenting process, without systematic decomposition or judgment \\[6pt]
& Empirical testing or provenance tracing present
& Analyze
& Systematically decomposing AI outputs through controlled variation or tracing output elements to their origins \\[6pt]
& Fluency--correctness distinction present
& Evaluate
& Judging the validity of system behavior against an independent criterion (correctness vs.\ surface fluency) \\
\midrule
\multirow{3}{2.2cm}{Final Proj.}
& Surface-level competencies
& Analyze
& Identifying components and relationships in the team's own system without specificity or cited evidence \\[6pt]
& Specific causal ethics reasoning or constraint specification
& Evaluate
& Making evidence-based judgments about downstream consequences; critiquing system behavior against defined standards \\[6pt]
& Specific failure mode design or integrated artifact construction
& Create
& Designing system components around identified limitations; producing a coherent, novel artifact integrating technical and ethical constraints \\
\bottomrule
\end{tabular}
\end{table}

As Fig.~\ref{fig:blooms-progression} shows, in HW1, reasoning was concentrated at the Remember/Understand and Apply levels (90\% combined), with students describing AI systems and referencing the information management pipeline but rarely decomposing, evaluating, or constructing. By Debate Round~1, one week later, the Apply and Analyze levels together accounted for 73\% of coded arguments, as the debate format required students to decompose claims through the pipeline rather than merely assert them. Evaluate-level reasoning appeared (12\%) but remained limited by the absence of architectural knowledge. The most substantial change occurred between Debate Round~1 and HW3, a span encompassing the deep learning modules (perceptron through transformer), the CNN studio and saliency interpretation assignment, and the midterm field experiment on chatbot reasoning. By HW3, the distribution had inverted relative to HW1: Analyze and Evaluate together accounted for 70\% of coded reasoning, with students empirically testing AI outputs, distinguishing fluency from correctness, and tracing provenance through training data. The Apply and Remember/Understand levels, which had dominated HW1, contracted to 25\%. By the final project, Create-level reasoning appeared for the first time at substantial scale (40\%), as teams designed AI-enabled systems with integrated safeguards, principled trade-offs, and constraint specifications. The combined Analyze, Evaluate, and Create share reached 95\%. The Remember/Understand level, which had accounted for 35\% of HW1 reasoning, was absent entirely.

This progression suggests that the course developed higher-order competencies through a deliberately-sequenced design in which each assessment built the competencies required by the next: the sector briefs supplied evidence for debates, the debates normalized the pipeline as an argumentative structure, the deep learning modules supplied architectural knowledge, the midterm's fluency-correctness finding transferred into independent creative evaluation, and the final project required integrating all prior competencies into a single artifact defended under external scrutiny. The upward trajectory evident in Fig.~\ref{fig:blooms-progression} is the aggregate result of that sequence.

\section{Course Adaptations}
\label{sec:adaptation}

The five mechanisms described in Section~\ref{sec:design} were developed together, and the reasoning progression documented in Section~\ref{sec:evidence} reflects their joint operation. Not every institution may be able to adopt the full design, for a variety of reasons. Therefore, this section identifies which mechanisms are separable, describes three potential adaptation pathways, and notes where partial adoption changes what can be expected.

\subsection{Separability of Mechanisms}

The five mechanisms differ in their adoptability.

\paragraph{Readily adoptable.} The conceptual pipeline (problem, data, approach, train/configure, metrics, evaluate, improve, reflect) is a design decision requiring no infrastructure beyond a syllabus. Any instructor can adopt it as the recurring structure of a course. Concurrent ethical integration is similarly portable: it requires threading ethical questions into technical assignments rather than segregating them into a dedicated module, which is a curricular choice rather than a resource dependency. Together, these two mechanisms establish the analytical vocabulary that the remaining three mechanisms then operationalize.

\paragraph{Adoptable with platform considerations.} The custom AI agent
was built within George Mason's PatriotAI platform, an institutional AI
gateway that provides a sandboxed environment with data protections
appropriate for educational use. The interaction template described in
Section~\ref{sec:agent} (scoped concept, scaffolded difficulty, active
assessment with formative feedback, student-controlled pacing) is not
platform-specific and so can be implemented on other systems that support
custom agent configuration. However, the instructor notes that platform choice is not neutral. Institutions adopting this mechanism must evaluate whether student interactions are retained or used for model training, whether access requires paid accounts that create access barriers, and whether the platform's behavior and terms remain stable across a semester.

\paragraph{Dependent on institutional conditions.} AI Studios require capped enrollment, a physical or synchronous meeting space, and an instructor comfortable with real-time facilitation, observation, and course correction during open-ended work sessions. The cumulative assessment portfolio requires a semester-length course with sufficient contact hours to support the full sequence from sector briefs through the final project. Both mechanisms are the primary sites where the reasoning progression documented in Section~\ref{sec:evidence} was observed to develop among students. Institutions that cannot provide these conditions may still adopt the above mechanisms but should not expect the same trajectory.

\subsection{Adaptation Pathways}

Three variants illustrate how the design scales to a different set of constraints.

\paragraph{General AI literacy without a technical progression.} An
institution seeking broad AI awareness rather than hands-on technical
practice can retain the conceptual pipeline, concurrent ethics
integration, one debate round, and the AI agent, while replacing the
deep learning modules and notebook-based assignments with guided
demonstrations and structured interaction with existing AI tools. This
variant preserves the analytical framework and the integration of
ethical and technical reasoning but removes the experimental components
(CNN ablations, midterm field experiment, artifact construction with
defended design rationale) that produced the Analyze-through-Create
progression in Fig.~\ref{fig:blooms-progression}. The proliferation of no-code
platforms has made artifact assembly broadly accessible, but assembling
an artifact is not equivalent to the Create level as defined in Bloom's
revised taxonomy. In the coding reported in Section~\ref{sec:evidence},
Create required students to design around identified failure modes,
specify constraints grounded in architectural understanding, and defend
trade-offs under external scrutiny. Without the technical progression
that supplies that understanding, artifact production remains at the
Apply level regardless of the tool's ease of use. The expected Bloom's
ceiling for this variant is Apply, with some students reaching Analyze
through the debate structure.

\paragraph{Discipline-embedded offerings.} The full scaffolding can be retained while replacing the cross-sector examples with discipline-specific content. For instance, in a healthcare variant, the sector briefs may analyze clinical AI deployments, the debates may address diagnostic AI and algorithmic triage, the midterm may probe medical chatbots, and the final project may build a health-related AI artifact. Analogous substitutions apply to policy (regulatory impact, surveillance, public resource allocation), business (demand forecasting, hiring algorithms, customer profiling), and the arts (generative tools, authorship, provenance). Because the pipeline, studios, and portfolio structure remain intact, the reasoning progression is expected to transfer, though the specific competencies developed will invariably reflect the disciplinary context.

\paragraph{Shorter format.} A one-credit or module-length offering can retain the conceptual pipeline, one debate round, and one hands-on assignment, collapsing the portfolio to two assessed components. The AI agent can supplement reduced contact hours. This variant sacrifices the cumulative feed-forward structure that distinguishes the full portfolio: without the midterm supplying the fluency-correctness finding, and without checkpoint-driven studios refining the final project, later assessments cannot build on competencies that earlier ones developed. The expected outcome is competency at the Apply level with partial access to Analyze.

\subsection{Dependencies and Expected Trade-offs}

The portable mechanisms (pipeline, concurrent ethics, agent template) establish a shared analytical vocabulary and ensure that ethical reasoning accompanies technical content from the outset. These are necessary conditions for the progression documented in this paper but are not sufficient to produce it. The reasoning trajectory from Remember/Understand to Create depended on the studios making methodological expectations explicit through protocols and real-time critique, and on the cumulative portfolio ensuring that each assessment supplied competencies the next one required. Removing either mechanism removes the conditions under which the documented progression was observed. Institutions (and instructors) adopting partial versions of the design should calibrate their expectations accordingly and, where possible, collect their own evidence of student reasoning to determine what their particular configuration supports.
\section{Conclusion}
\label{sec:conclusion}

This paper described the design, assessment structure, and formative evidence 
from UNIV~182, a semester-long AI literacy course at George Mason University 
in which undergraduates across majors learn to understand, use, evaluate, and 
build AI systems without prior technical preparation. The course is organized 
around five interdependent mechanisms: a conceptual pipeline traversed at 
increasing levels of technical sophistication, integration of ethical reasoning 
with technical content throughout the semester, AI Studios for structured 
in-class practice, a cumulative assessment portfolio in which each assignment 
builds competencies required by the next, and a custom AI agent for 
out-of-class reinforcement. A comparative taxonomy of cross-major AI literacy 
efforts situates the course within the existing landscape.

Instructor-coded analysis of student artifacts across four assessment stages 
documented a progression from descriptive, intuition-based reasoning toward 
technically grounded design with integrated safeguards. In HW1, 90\% of coded 
reasoning operated at the Remember/Understand and Apply levels. By the final 
project, 95\% operated at the Analyze, Evaluate, or Create levels, with Create 
appearing at 40\% for the first time. The midterm field experiment, in which 
student teams designed reasoning probes for consumer chatbots, was 
subsequently developed with students into a co-authored research 
paper~\cite{shehu2025consumerchatbotsreasonstudentled}, providing additional 
evidence that non-major undergraduates can contribute to publishable AI 
research when the course structure supports the progression from structured 
inquiry to independent analysis.

The repeated traversal of the conceptual pipeline supported cumulative 
learning: students encountered the pipeline first through spam filtering, then 
through CNN architectures, then through ethically consequential deployments, 
and each pass increased in sophistication while retaining the same structure. 
By the final project, students addressed technical feasibility and ethical 
considerations together rather than separately, consistent with the integrated 
design of the course. The AI Studios made methodological expectations explicit 
through protocols (plan before beginning, log decisions, review in real time), 
and students applied those practices in subsequent independent work. The 
progression from structured studio sessions to public defense before external 
evaluators functioned as a transfer test: early work occurred with instructor 
scaffolding, and later work required defending choices under questioning from 
people with no prior familiarity with the projects. Section~\ref{sec:adaptation} 
identified which of these mechanisms are separable, and how the design can be 
adapted for contexts ranging from general AI literacy without a technical 
progression to discipline-embedded offerings in healthcare, policy, business, 
and the arts.

Several limitations warrant explicit acknowledgment. The students who completed the course may 
have been self-selected for motivation or comfort with sustained technical 
challenge; the current design cannot distinguish scaffolding effects from 
selection effects. Three operational areas required ongoing adjustment: 
timeboxing studio activities, where the open-ended nature of exploratory work 
required real-time instructor judgment that raises questions about portability 
to larger sections or less-experienced instructors; the transition from no-code 
tools to light coding, which was smoother for some students than others and may 
benefit from an additional bridging activity; and team dynamics in the final 
project, which varied in quality and may be improved through structured team 
formation protocols. The course was designed and taught by a single instructor. 
Work in~\cite{henderson2011facilitating} identifies instructor 
dependence as a central challenge in STEM education reform, and the protocols, 
rubrics, and structured checkpoints described in this paper are intended to 
make as much of the instructor's judgment explicit as possible without claiming 
that a different instructor would produce the same outcomes.

Several questions remain open, including whether the competencies developed in this course 
persist beyond the semester, and whether the design transfers across institutional 
types with different student populations and resource constraints. With these acknowledged, this course is offered as a documented example for educators and institutions seeking to prepare students across disciplines to engage with AI systems as informed, technically-capable, and ethically-grounded participants.

% --------------------------------------------------------------------
% Acknowledgments
% --------------------------------------------------------------------
\section*{Acknowledgments}

To the UNIV 182 Fall 2025 students: You arrived uncertain about what AI literacy meant and left as builders, critics, and advocates. Your willingness to push through technically-rigorous concepts, debate seriously and generously, build what mattered to you, and defend your work openly made this course what it became. Several of you told me this changed how you see your own capabilities; it changed how I see what is possible with students across all majors. To the Spring 2026 students: keep working towards that finish line; you will find it rewarding. Thank you all for your trust, your curiosity, and your commitment to the work.

% ---------- References ----------
\bibliographystyle{unsrtnat} % change if you prefer a different style
\bibliography{references}

\end{document}